\documentclass[%
reprint,
superscriptaddress,
showpacs,preprintnumbers,
nofootinbib,
 amsmath,amssymb,
 aps,
prb,
]{revtex4-1}

\usepackage{graphicx}
\usepackage{dcolumn}
\usepackage{bm}

\begin{document}


\title{Diagonal superexchange in a simple square CuO$_2$ lattice}


\author{V. A. Gavrichkov}
\affiliation{Kirensky Institute of Physics, Siberian Branch of the Russian Academy of Sciences, 660036 Krasnoyarsk, Russia}
\affiliation{Rome International Center for Materials Science Superstripes RICMASS, via del Sabelli 119A, 00185, Roma, Italy}
\author{S. I. Polukeev}
\affiliation{Kirensky Institute of Physics, Siberian Branch of the Russian Academy of Sciences, 660036 Krasnoyarsk, Russia}
\affiliation{Siberian Federal University, Svobodny Prospekt 79, Krasnoyarsk, 660041, Russia}
\author{S. G. Ovchinnikov}
\affiliation{Kirensky Institute of Physics, Siberian Branch of the Russian Academy of Sciences, 660036 Krasnoyarsk, Russia}

\date{\today}

\begin{abstract}

Many microscopic models with the interaction between the next-nearest neighbours as a key parameter for cuprate physics have inspired us to study the diagonal superexchange interaction in a CuO$_2$ layer. Our investigation shows that models with extended hopping provide a correct representation of magnetic interactions only in a hypothetical square CuO$_2$ layer, where the diagonal superexchange interaction with the next-nearest neighbors always has the AFM nature. The conclusions are based on the symmetry prohibition on FM contribution to the diagonal superexchange between the next-nearest neighbors for a simple square CuO$_2$ layer rather than for a real CuO$_2$ layer, where diagonal AFM superexchange may be overestimated. We also discuss the reasons for magnetic frustration effects and high sensitivity of spin nanoinhomogeneity to square symmetry breaking.
\end{abstract}

\maketitle


\section{\label{sec:intr}Introduction\\}

The standard spin wave theory describes well some, but not all of the properties of the magnon spectrum in antiferromagnetic (AFM) cuprates~\cite{Annet_etal1989, Kim1989, Coldea2001, Katanin_2002, Toader_2005, Headings_2010, Dean_2012, Yamamoto_2019, Bao_2025}, and  failing to capture the continuum observed at high energies~\cite{Bao_2025}. For this reason, it is important to understand whether the superexchange theory based on the single-band Hubbard model and currently used to understand magnetism in parent cuprates provides a reliable description of the experiment picture. Alternatively, the single-band Hubbard model itself might need modification. The simplest extension would be hopping $t'$ on next-nearest neighbors, a parameter which has been argued to play an important role in superconductivity~\cite{Qin2020, Xu_2024}, and calculating beyond the low-energy approximation in a $pd$ model as the initial microscopic model for the Hubbard approach ~\cite{Feiner_etal1996, Raimondy_etal1996, Gavrichkov_2000}. The extension can be done in the approach that was previously used to study pressure and optical pumping effects on the superexhange interaction in transition metal oxides~\cite{Gavrichkov2017, Gavrichkov2016, Gavrichkov2020, Mikhaylovskiy2020}. However, these possibilities remained beyond the previous study, despite being discussed for a long time~\cite{Jiang2019, White1999, Chung2020, Qin2020, Wei2024}.

Indeed, the parameters: hopping $t'$ and superexchange with next-nearest neighbors are relevant in a number of approaches to studying the physics of high-$T_c$ superconducting (HTSC) cuprates. The parameters could be important not only for the extended hopping problem in a set of $t$-$J$, $t$-$t'$-$J$, $t$-$t'$-$t''$-$J$   models with a realistic hole-pairing mechanism in HTSC cuprates  (see ~\cite{Jiang2022} and references therein), but also for understanding the nature of a pseudogap state with  $\vec k$ arcs in the antinodal direction of the Fermi surface in  $\vec k$-dependent experiments~\cite{Hartstein2020a, Kanigel2006}. Moreover, it is known that the temperature window for the pseudogap state shrinks with the increasing next-nearest neighbor hopping, which indicates that diagonal hopping may not be supportive of the pseudogap features ~\cite{AlRashid2024}. The results of dynamic cluster quantum Monte-Carlo simulations for the Lifshitz transition of the two-dimensional(2D) Hubbard model show the sensitivity to the magnitude of diagonal next-nearest neighbor hopping to be a control parameter ~\cite{Chen2012}. Complicated spectral features of the two-dimensional Hubbard model are simply interpreted near the Mott transition by considering how the next-nearest neighbor hopping shifts spectral weights~\cite{Kohno2014}.   The momentum-sector-selective metal-insulator transitions in the eight-site dynamical cluster approximation for the 2D Hubbard model are explored on a phase diagram in the space of interaction and second-neighbor hopping control parameters ~\cite{Gull2009}.
The magnitude of the interaction of diagonal neighbors plays a key role in the study of spin (SDW) and charge(CDW) nanoinhomogeneity in cuprate materials, where both observed phases have tilted stripes (the so called  "Y shift") with the same degree of tilting ~\cite{Wei2024}, and the origin of tilting can be explained by a small anisotropy in hopping between the next-near neighbors. The specific alignment direction of the stripes is highly sensitive to the hopping interaction where even a small anisotropy in it can result in subtle observed tilting in LSCO ~\cite{Jiang2019, Wakimoto1999, Fujita2002} and LBCO ~\cite{Dunsiger2008} samples.

However, the studies of the effects of next-nearest neighbor hopping and magnetic frustrations on the spectrum of quasiparticles are still based on assumptions regarding the sign and magnitude of hopping and superexchange interaction $J_{tot} \left( {\vec R_{11} } \right)$ between the diagonal neighbors. Previous calculations provide contradictory results regarding the nature of the superexchange interaction $J_{tot} \left( {\vec R_{11} } \right)$, both AFM~\cite{Annet_etal1989, Moreira2006} and FM ~\cite{Wan2009} type.
It is certainly not enough to understand the whole range of phenomena associated with these parameters. Before we derived a simple rule for detecting the sign of contribution to the superexchange interaction from a single virtual electron-hole pair~\cite{Gavrichkov2017, Gavrichkov2020}. Our conclusions concerned only the interaction with the nearest neighboring magnetic  ions. For the CuO$_2$ layer of parent cuprates, this magnitude $J_{tot} \left( {\vec R_{01} } \right) \approx 0.15eV$ ~\cite{Gavrichkov2016} is in agreement with the neutron scattering data ~\cite{Coldea2001}, and the approach itself allows us to study the dependence of superexchange in $3d$ oxides on external factors: applied pressure~\cite{Gavrichkov2016, Gavrichkov2020} and optical pumping ~\cite{Mikhaylovskiy2020}.

 \begin{table*}
 \caption{Notations and Specification  \label{tab:1}}
 \begin{ruledtabular}
 \begin{tabular}{@{}ll@{}}
  \footnotesize Notations             & \footnotesize Specification \\
\hline \\
  \footnotesize $\Delta _{nS}  = E_{nS}  + E_{{}^1A}  - 2\varepsilon _{b_{1g} }$ & \footnotesize It is a simple analogue of the well-known Hubbard repulsion $U$ in the Hubbard multiband  model. \\
  \footnotesize  $\Delta _{mT}  = E_{mT}  + E_{{}^1A}  - 2\varepsilon _{b_{1g} } $ & \footnotesize This is clearly seen if we take into account that $U$  is the strongest interaction in $\Delta _{nS}$ and \\
  \footnotesize     & \footnotesize $\Delta _{nS}  \approx U\left( {C_2^{N_\lambda + 1}  + C_2^{N_\lambda - 1}  - 2C_2^{N_\lambda} } \right) = U$, where $C_2^{N_\lambda}  = \frac{{N_\lambda!}}{{\left( {N_\lambda - 2} \right)!2}}$  is a number of combinations of  $N_\lambda$ \\

   \footnotesize              & \footnotesize by 2. Similarly for $\Delta _{mT}  = E_{mT}  + E_{{}^1A}  - 2\varepsilon _{b_{1g} }  \approx U$ , and the differences between $\Delta _{mT} $  and $
\Delta _{nS}$ are \\
\footnotesize & \footnotesize   associated with the Hund exchange $J_H \approx 1eV$  and splitting of terms of different symmetry in the crystal \\
\footnotesize             & \footnotesize field  $\delta _ \bot   \approx \varepsilon _{b_{1g} }  - \varepsilon _{a_{1g} }=2 eV$    \\
\\ \hline \\
\footnotesize $t_{\sigma}^{r_0 ,r_{nS} } \left( {\vec R_{ij}} \right)$             & \footnotesize - hopping integrals where  $r_0  = \left( {{}^1A,b^s_{1g} } \right)$ and $r_{nS}  = \left( {b^s_{1g} ,nS} \right)$, $r_{mT}  = \left( {b^s_{1g} ,mT} \right)$  are the root vectors for \\
 \footnotesize $t_{\sigma}^{r_0 ,r_{mT} } \left( {\vec R_{ij}} \right)$ & \footnotesize  quasiparticles with initial and final states  in the configuration space $N_ +  \left( {d^{10} } \right),N_0 \left( {d^9 } \right),N_ -  \left( {d^8 } \right)$ in Fig.2a \\
\\
\hline \\
\footnotesize $t_{\sigma}^{r_0 ,r_{mT} } \left( {\vec R_{ij} } \right) = $
 & \footnotesize - the hopping integral which in the Hubbard operator representation corresponds to the quasiparticle   \\
 \footnotesize $\sum\limits_{\lambda \lambda '} {t_{}^{\lambda \lambda '} \left( {\vec R_{ij} } \right)\gamma _{\lambda \sigma }^* (r_0 )\gamma _{\lambda '\sigma } (r_{mT} )} $
 & \footnotesize transfer as a lattice sequence of intra-cell transitions between the multielectron cell states\\
 \hline \\
 \footnotesize $S_ +   = S_ -  $      & \footnotesize - FM condition where $S_ +   = 0$
  spin in the $\left| {{}^1A} \right\rangle $ state in the sector $N_ +  \left( {d^{10} } \right)$ and $S_ -   = 0,1$ spin in  \\
 \footnotesize             & \footnotesize $\left| {h = nS,mT} \right\rangle $ states in the sector $N_ -  \left( {d^8 } \right)$ (see Fig.2) \\
 \\ \hline \\
 \footnotesize $S_ +   = S_ -   \pm 1$ & \footnotesize - AFM condition where $S_ +  $ spin in the $\left| {{}^1A} \right\rangle $ state in the sector $N_ +  \left( {d^{10} } \right)$ and  $S_ -   = 0,1$
 spin in  \\
 \footnotesize        & \footnotesize $\left| {h = nS,mT} \right\rangle $ states in the sector $N_ -  \left( {d^8 } \right)$ (see Fig.2) \\
\\
 \end{tabular}
 \end{ruledtabular}
 \end{table*}

In parent cuprates, virtual hopping and electron-hole pairs on the (next-)nearest neighbors can also lead to superexchange between them. Here, we show the superexchange  constant $J_{tot} \left( {\vec R_{ij} } \right)$, important for other approaches, to behave in an unusual way. Indeed, the contribution from the virtual electron-hole pair with the ${}B_{1g}$ hole symmetry, where the   ${}^3B_{1g}$ triplet band competes in energy with the Zhang-Rice singlet ${}^1A_{1g}$ band ~\cite{Kamimura_etal1990, Eto1991, Raimondy_etal1996, Janowitz2004, Gavrichkov2016} has a zero magnitude at the diagonal directions in the square CuO$_2$ layer. This leads exactly to a zero FM contribution to the total magnetic interaction  $J_{tot} \left( {\vec R_{11} } \right)$ between the next-nearest neighbors of  Cu$^{2+}$  ions. As a consequence, the latter has  purely AFM nature $J_{tot} \left( {\vec R_{11} } \right) < 0$, but it quickly decreases with increasing distance between the Cu$^{2+}$  ions.

The paper is organized as follows: in the next section (Sec.\ref{sec:II}) we provide theoretical background on the many-electron approach to the study of superexchange interactions in parent cuprates.
Symmetric effects on superexchange in the CuO$_2$ layer with the square symmetry are given in detail in Sec.\ref{sec:III}. Discussion and conclusion are presented in Sec.\ref{sec:IV}. Details of the calculations are taken out into Appendix A and B.

\section{\label{sec:II}Hamiltonian}

\begin{figure}
\includegraphics{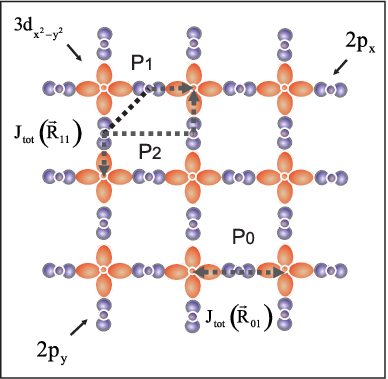}
\caption{Paths P$_0$, P$_1$  and P$_2$ of the superexchange interactions $J_{tot} \left( {\vec R_{01} } \right)$ and $J_{tot} \left( {\vec R_{11} } \right)$ for the first and the second neighbor  ions with the participation of $2p$ oxygen orbitals forming  $\sigma$  overlapping with $3d$ copper ions, and $90^\circ$(P$_1$) or small $\pi$(P$_2$)   - overlapping between themselves. Here, the interactions  $J_{tot} \left( {\vec R_{01} } \right)$ are of  AFM nature, but the magnitude of  the  $J_{tot} \left( {\vec R_{11} } \right)$  interaction with the second  neighbors  is  still  unknown.}
\label{fig:1}
\end{figure}

In this section, we investigate the sign and magnitude of the superexchange interaction $J_{tot} \left( {\vec R_{11} } \right)$ with next-nearest-neighbor Cu$^{2+}$  ions through different oxygen orbitals in the CuO$_2$ layer (see Fig.\ref{fig:1}). Indeed, in the magnetic interaction with the second neighbors, the overlapping oxygen orbitals play a significant role.
There are two paths P$_1$ and P$_2$  to create various virtual electron-hole pairs. They are distinguished by the overlapping oxygen orbitals. There are $90^\circ$ degree overlapping oxygen orbitals in the path P$_1$, and small $\pi$ - overlapping in the path P$_2$. The virtual electron-hole pairs in both paths can generate both AFM and FM contributions to the superexchange interaction $J_{tot} \left( {\vec R_{11} } \right)$ between the diagonal second neighbor  Cu$^{2+}$ ions.

\begin{figure*}
\includegraphics{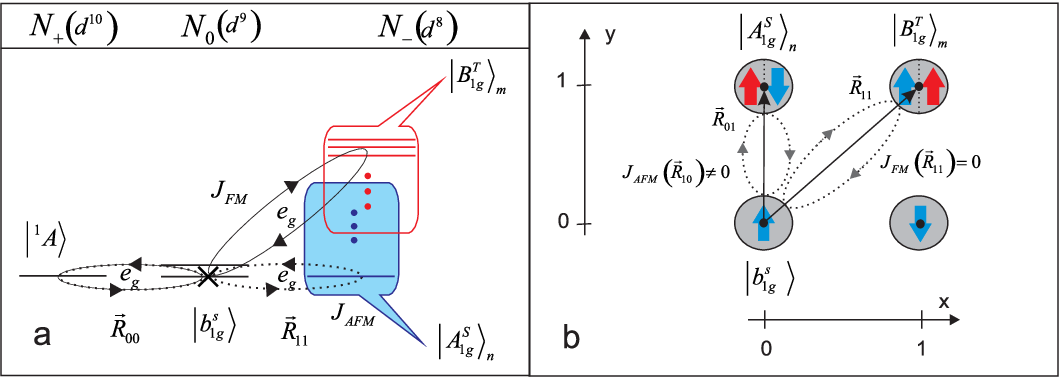}
\caption{(a): A configuration space of the unit cell of CuO$_2$ layer. The cross denotes the occupied hole eigenstates $\left| {b^s_{1g} } \right\rangle$ in the $N_0 \left( {d^9 } \right)$ sector. Ellipses correspond to the virtual $e_g$ electron-hole pairs with the $J_{FM}^{} \left( {\vec R_{ij} } \right)$ and  $J_{AFM} \left( {\vec R_{ij} } \right)$
contributions to the total exchange interaction $\hat H_S$. (b):  A lattice diagram  of the direct and diagonal superexchange interactions in the cell representation Eq.(\ref{eq:4}) for the square  CuO$_2$ layer.}
\label{fig:2}
\end{figure*}

The superexchange constant $J_{tot} \left( {\vec R_{ij} } \right)$ in the Eq.(\ref{eq:1}) is additive over all possible states in the electronic $N_ +  \left( {\left| {A_1 } \right\rangle } \right)$ and two-hole  $N_ -  \left( {\left| {{}^1A_{1g} } \right\rangle _{nS} ,\left| {{}^3B_{1g} } \right\rangle _{mT} ,...} \right)$
sectors in Fig.\ref{fig:2}, and the superexchange interaction (\ref{eq:1}) is obtained in the second order of the cell perturbation theory over interband contributions  to the total Hamiltonian  $\hat H$ from the interatomic hopping contribution ~\cite{Anderson1959, Chao_etal1977, Gavrichkov2017}.
\begin{equation}
\hat H_S  = \hat H_{AFM}  + \hat H_{FM},
\label{eq:1}
\end{equation}
where $\hat H_{AFM}  = \sum\limits_{ij} {J_{AFM}^{} \left( {\vec R_{ij} } \right)} \left( {\vec s_i \vec s_j  - {\textstyle{1 \over 4}}n_i n_j } \right)$, $\hat H_{FM}  = \sum\limits_{ij} {J_{FM}^{} \left( {\vec R_{ij} } \right)} \left( {\vec s_i \vec s_j  + {\textstyle{3 \over 4}}n_i n_j } \right)$, and the exchange constants $J_{AFM}^{} \left( {\vec R_{ij} } \right)>0$, $J_{FM}^{} \left( {\vec R_{ij} } \right)<0$ are given by

\begin{widetext}
\begin{eqnarray}
J_{AFM} \left( {\vec R_{ij} } \right) &=& \sum\limits_{nS} {J_{AFM}^{\left( nS \right)} } (\vec R_{ij} ) = \sum\limits_{nS = 1}^{N_S } {{{\left| {t_{}^{r_0,r_{nS}} \left( {R_{ij} } \right)} \right|^2 } \mathord{\left/
 {\vphantom {{\left| {t_{}^{r_0,r_{nS}} \left( {R_{ij} } \right)} \right|^2 } {\Delta _{nS} }}} \right.
 \kern-\nulldelimiterspace} {\Delta _{nS} }}} ,\;\;\Delta _{nS}  = E_{nS}  + E_{{}^1A}  - 2\varepsilon _{b_{1g} } \label{eq:2}  \\
 J_{FM}^{} \left( {\vec R_{ij} } \right) &=& \sum\limits_{mT} {J_{FM}^{\left( mT \right)} \left( {\vec R_{ij} } \right)}  =  - \sum\limits_{mT = 1}^{N_T } {{{\left| {t_{}^{r_0,r_{mT}} \left( {Rij} \right)} \right|^2 } \mathord{\left/
 {\vphantom {{\left| {t_{}^{r_0,r_{mT}} \left( {Rij} \right)} \right|^2 } {2\Delta _{mT} }}} \right.
 \kern-\nulldelimiterspace} {2\Delta _{mT} }}} ,\;\;\Delta _{mT}  = E_{mT}  + E_{{}^1A}  - 2\varepsilon _{b_{1g} }.
\nonumber
\end{eqnarray}
\end{widetext}

where the indices $nS$ and $mT$ with $T=0,2\sigma$  run over the spin singlets from 1 to $N_S$ and over the spin triplets from 1 to   $N_T$. Since the derivation of the spin Hamiltonian (\ref{eq:1}) can be found in the work ~\cite{Gavrichkov2017}, we only provide a brief discussion of the one in Appendices A and B.

We have also included all the notations used here and below into a Table 1, for example the generalized Hubbard repulsions $\Delta _{nS}$,  $\Delta _{mT}$  and hopping integrals $t_{}^{r_0 ,r_{nS}} \left( {\vec R} \right)$, $t_{}^{r_0 ,r_{mT}} \left( {\vec R} \right)$ used in the Eq.(\ref{eq:2}).

 Virtual electron-hole excitations through the dielectric gap  $\Delta _h  = E_h  + E_{{}^1A}  - 2\varepsilon _{b_{1g} }$ to the conduction band and vice versa in the Eq.(\ref{eq:2}) contribute to the superexchange interaction $\hat H_S $  ~\cite{Gavrichkov2017}:
\begin{eqnarray}
t_{\sigma}^{0,nS} &\left( {\vec R_{ij} } \right)& \equiv t_{}^{\left( {{}^1Ab_{1g}^{s} } \right),\left( {b_{1g}^{s} nS} \right)} \left( {\vec R_{ij} } \right) = \label{eq:3} \\
&=& \sum\limits_{\lambda \lambda '} {t_{\lambda \lambda '} \left( {\vec R_{ij} } \right)\gamma _{\lambda \sigma }^* ({}^1Ab_{1g}^{\sigma} )\gamma _{\lambda '\sigma } (b_{1g}^{\sigma} nS)} \nonumber \\
t_{\sigma}^{0,mT} &\left( {\vec R_{ij} } \right)& \equiv t_{}^{\left( {{}^1Ab_{1g}^{s}} \right),\left( {b_{1g}^{s} mT} \right)} \left( {\vec R_{ij} } \right) = \nonumber \\
&=& \sum\limits_{\lambda \lambda '} {t_{\lambda \lambda '} \left( {\vec R_{ij} } \right)\gamma _{\lambda \sigma }^* ({}^1Ab_{1g}^{\sigma})\gamma _{\lambda '\sigma } (b_{1g}^{\sigma} mT)}.
\nonumber
\end{eqnarray}
The spin Hamiltonian (\ref{eq:1}) was derived from the initial $pd$ Hamiltonian (see Eq.(\ref{eq:A.1}) in the Appendix A) by using the projective operator method ~\cite{Chao_etal1977, Gavrichkov2017}(see Appendix B). For details of deriving the multi-electron Hamiltonian (\ref{eq:A.12})  from the $pd$ model, see the Appendix A and works~\cite{Feiner_etal1996,  Raimondy_etal1996, Gavrichkov_2000}, where a five-orbital basis $p_\lambda  ,\left( {\lambda  = x,y,z} \right),d_{x^2  - y^2 }(d_x) ,d_{z^2}(d_z)$ is typically used.
The initial copper $Cu^{2+}$  ion is in a state with the $\left| {2b^s_{1g} } \right\rangle$ hole at the lowest $\varepsilon _{b_{1g} }$ energy in the $N_0(d^9)$ sector (Fig.\ref{fig:2}). By rearranging the additional virtual hole over the oxygen $b_{1g}$ and $a_{1g}$, $p_z$(apical oxygen) and  copper $d_x$, $d_z$ - orbitals, we obtain $9$ of the $^1A_{1g}$ -, $6$ of the $^1B_{1g}$ - singlets and $4$ of the $^3A_{1g}$ -, $6$ of the $^3B_{1g}$ - triplets with energies $E_{nS}$ and $E_{mT}$   respectively. All the possible eigenstates $\left| {{}^1A_{1g} } \right\rangle _{nS}$ and $\left| {{}^3B_{1g} } \right\rangle _{mT}$ in the configuration space in the Fig.\ref{fig:2}, as well as their energies $E_h$ with $h = nS,mT$ are obtained in the exact diagonalization procedure for the intra-cell part  of the multi-orbital $pd$ model ~\cite{Gavrichkov_2000}(see also Appendix A).

The sum of all the FM and AFM contributions corresponds to the sum over all possible virtual electron-hole pairs (so called exchange loops ~\cite{Gavrichkov2020}). The FM or AFM nature of the contribution from a specific exchange loop is easily determined using the rule:  if $S_ +   = S_ -$   there is an AFM contribution, in the case of $S_ +   = S_ -   \pm 1$ there is an FM contribution, where $S_ \pm$ is the spin of the electron and hole in the specific virtual pair ~\cite{Irkhin1994, Gavrichkov2020}. All the triplet states  $\left| {mT} \right\rangle$ in the hole sector $N_ -  \left( {d^8 } \right)$ contribute $J_{FM}^{} \left( {\vec R_{ij} } \right)$, and all the singlet states $\left| {nS} \right\rangle$ contribute $J_{AFM} \left( {\vec R_{ij} } \right)$ to the exchange constant $J_{tot} \left( {\vec R_{ij} } \right) $. Thus, the dependence of the superexchange interaction on the distance between the interacting Cu$^{2+}$ ions can be calculated using $t_{}^{0,ns} \left( {\vec R_{ij} } \right)$ and $t_{}^{0,mT} \left( {\vec R_{ij} } \right)$  hopping integrals in the Eq.(\ref{eq:2}).

\section{\label{sec:III} Diagonal superexchange interaction in the square symmetry of the $CuO_2$ layer}

Where are the effects of the CuO$_2$ layer symmetry hidden in the calculation of the $J_{tot} \left( {R_{11} } \right)$ exchange constant using Eq.(\ref{eq:2})? Since in the Mott-Hubbard materials the charge transfer is considered as a lattice sequence of intraionic transitions,~\cite{Hubbard_1963} we can expect symmetry effects in this case. To begin with, there is the point $C_4$ symmetry  of the CuO$_6$ octahedron in the procedure of exact diagonalization of the intra-cell part of the Hamiltonian of the $pd$ model ~\cite{Gavrichkov_2000}. Indeed,  there are $C_{2N_\lambda  }^2  = N_S  + 3N_T $ of the spin singlets $N_S  = C_{N_\lambda  }^2  + N_\lambda  $ and triplets $N_T  = C_{N_\lambda  }^2 $ in the two-hole  $N_ -  \left( {d^8 } \right)$ sector (Fig.\ref{fig:2}) within the  $N_\lambda  $ orbital approach, where  $C_n^k $ is the number of combinations. For example, in the five-orbital approach there are 15  AFM ($N_S  = 15$ ) and 10 FM ($N_T  = 10$ ) contributions to the total superexchange interaction   $J_{tot} \left( {\vec R_{ij} } \right)$ from various virtual electron-hole pairs.

Using the intra-cell part of the multiorbital $pd$ Hamiltonian in the symmetric representation of canonical fermions ~\cite{Shastry_1989}, all the spin singlet states $\left| {{}^1A_{1g} } \right\rangle _{nS}$ and $\left| {{}^1B_{1g} } \right\rangle _{mT}$, spin triplet states $\left| {{}^3A_{1g} } \right\rangle _{nS}$ and $\left| {{}^3B_{1g} } \right\rangle _{mT}$, and also single hole spin doublet states $\left| {a^s_{1g} } \right\rangle$,  $\left| {b^s_{1g} } \right\rangle$ can be obtained in the exact diagonalization procedure for the eigenvalue problem in different sectors: $N_ -  \left( {d^8 } \right)$, $N_0 \left( {d^9 } \right)$, $N_ +  \left( {d^{10} } \right)$ of the configuration space (see Appendix A). To solve a problem  associated with taking into account the common oxygen ion in the CuO$_2$ layer, the initial $pd$  Hamiltonian was rewritten in the representation of symmetrized Bloch states of oxygen $2p$ ions~\cite{Shastry_1989, Feiner_etal1996, Feiner_etalPRL1996, Gavrichkov_2000}:

\begin{widetext}
\begin{eqnarray}
\left( {\begin{array}{*{20}c}
   {b_{\vec k\sigma } }  \\
   {a_{\vec k\sigma } }  \\
\end{array}} \right) = \hat {\rm P}\left( {k_x ,k_y } \right)\left( {\begin{array}{*{20}c}
   {p_{x\vec k\sigma } }  \\
   {p_{y\vec k\sigma } }  \\
\end{array}} \right) = {\raise0.7ex\hbox{$i$} \!\mathord{\left/
 {\vphantom {i {\mu _{\vec k} }}}\right.\kern-\nulldelimiterspace}
\!\lower0.7ex\hbox{${\mu _{\vec k} }$}}\left( {\begin{array}{*{20}c}
   {s_x \left( {\vec k} \right)} & {s_y \left( {\vec k} \right)}  \\
   {{\mathop{\rm sgn}} \left( {k_x k_y } \right)s_y \left( {\vec k} \right)} & { - {\mathop{\rm sgn}} \left( {k_x k_y } \right)s_x \left( {\vec k} \right)}  \\
\end{array}} \right)\left( {\begin{array}{*{20}c}
   {p_{x\vec k\sigma } }  \\
   {p_{y\vec k\sigma } }  \\
\end{array}} \right),
\label{eq:4}
\end{eqnarray}
\end{widetext}
where ${\hat P^2\left( {k_x ,k_y } \right)}  = 1$, and the coefficients $\mu _{\vec k}  = \sqrt {s_x^2 \left( {\vec k} \right) + s_y^2 \left( {\vec k} \right)}$ with $s_x \left( {\vec k} \right) = \sin \left( {{{k_x } \mathord{\left/
 {\vphantom {{k_x } 2}} \right.
 \kern-\nulldelimiterspace} 2}} \right)$ and $s_y \left( {\vec k} \right) = \sin \left( {{{k_y } \mathord{\left/
 {\vphantom {{k_y } 2}} \right.
 \kern-\nulldelimiterspace} 2}} \right)$ constructed on the square lattice of the CuO$_2$ layer . As a consequence, the initial $pd$ Hamiltonian (\ref{eq:A.1}) in the cell representation of symmetrized Wannier states can be renormalized by using the coefficients $\lambda _{\vec k}  = \frac{{2s_x s_y }}{{\mu _{\vec k} }}$,  $\xi _{\vec k}  = \frac{{s_x^2  - s_y^2 }}{{\mu _{\vec k} }}$,  for $pd$ hopping and $\nu _{\vec k}  = \frac{{2s_x^2 s_y^2 }}{{\mu _{\vec k}^2 }}$,  $\chi _{\vec k}  = \frac{{2s_x s_y }}{{\mu _{\vec k}^2 }}\left( {s_x^2  - s_y^2 } \right)$ for $pp$ hopping, two of which $\xi \left( {\vec R_{ij} } \right) = \frac{1}{{\sqrt N }}\sum\limits_{\vec k} {\xi _{\vec k} }$ and $\chi \left( {\vec R_{ij} } \right)$ are equal to zero for diagonal hopping with  $\vec R_{ij}  = \vec R_{11}$. Indeed, the square lattice remains invariant upon the replacement $x\leftrightarrows y$.

\begin{figure*}
\includegraphics{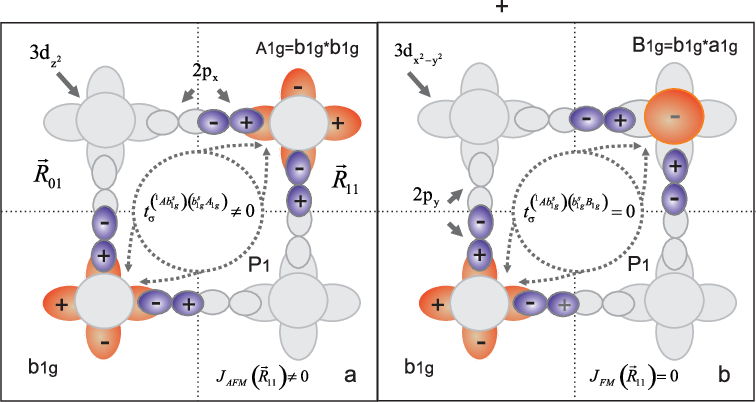}
\caption{ Diagram of the CuO$_2$ layer in the symmetry cell representation of the $a_{1g}$ and $b_{1g}$ oxygen  orbitals (see Eq.\ref{eq:4}) for diagonal: (a) AFM and (b) FM superexchange interactions. The color indicates the orbitals involved into the virtual electron-hole pairs at the next-neighboring copper ions. On the right side (b) it is clearly seen that due to the zero diagonal $pp$ and $pd$ overlapping, the FM contribution from a virtual pair with $B_{1g}$  symmetry is impossible.}
\label{fig:3}
\end{figure*}

The symmetry of the virtual electron-hole pair in Fig.\ref{fig:3}  is determined by the symmetry of the $\left| {A_{1g} } \right\rangle_n$ and $\left| {B^T_{1g} } \right\rangle_m$ two-hole states, since the virtual electron can only be in the state $\left| {{}^1A} \right\rangle$ of a completely occupied shell  in the sector $N_+$ (Fig.\ref{fig:2}). The coefficients  $\xi _{\vec k}$, $\chi _{\vec k}$ renormalize only the contributions with the holes in the $\left| {B^T_{1g} } \right\rangle_m$  states, in the hopping Hamiltonian $\hat H_{cc}$

\begin{widetext}
\begin{eqnarray}
t_{\sigma}^{0,B_{1g}}(R_{ij})&=&\frac{2t_{pd_x}}{\sqrt{3}}\xi_{ij}  {\gamma^*_{b\sigma}(^1A,b^{s}_{1g})\gamma_{d_z\sigma}(b^{s}_{1g},B_{1g})}+2t_{pp} \chi_{ij} {\gamma^*_{b\sigma}(^1A,b^{s}_{1g})\gamma_{a\sigma}(b^{s}_{1g},B_{1g})} \label{eq:5} \\
&-&2t^{(ap)}_{pp}\xi_{ij}\sum\limits_{ss'}{\gamma^*_{b\sigma}(^1A,b^{s}_{1g})\gamma_{p_z}(b^{s}_{1g},B^{T}_{1g})}
 = \left\{ {\begin{array}{*{20}c}
   {0,i = j}  \\
   {\emptyset,i \ne j}  \\
\end{array}} \right., \nonumber \\
t_{\sigma}^{0,A_{1g}}(R_{ij})&=&-2t_{pd_x}\mu_{ij}  {\gamma^*_{b\sigma}(^1A,b^{s}_{1g})\gamma_{d_x\sigma}(b^{s}_{1g},A_{1g})}-2t_{pp} \chi_{ij} {\gamma^*_{b\sigma}(^1A,b^{s}_{1g})\gamma_{b\sigma}(b^{s}_{1g},A_{1g})}\neq 0 \label{eq:6}
\end{eqnarray}
\end{widetext}
at any pairs of indices $\sigma$,$s$ and $n$,$m$, where the latter are not shown. The hopping processes $\hat{h}^{(a)}$ in the Eq.(\ref{eq:A.2}) does not contribute to the superexchange since the magnetic cell is in the $|b^s_{1g}\rangle$ state. Therefore, it is better to group the partial contributions to the total superexchange interaction $\hat H_S $ not by their singlet or triplet spin nature, but by the orbital symmetry of the virtual electron-hole pair, which can be in different orbital states with the  $A_{1g}$ and  $B_{1g}$ in $C_4$ point symmetry (see Fig.\ref{fig:3}). Thus, instead of the Eq.(\ref{eq:2}) we obtain
\begin{equation}
J_{tot} \left( {\vec R_{ij} } \right) = \Delta J_{AFM}^{A_{1g} } \left( {\vec R_{ij} } \right) + \Delta J_{FM}^{B_{1g} } \left( {\vec R_{ij} } \right),
\label{eq:7}
\end{equation}
where
\begin{eqnarray}
\Delta J_{AFM}^{A_{1g} } \left( {\vec R_{ij} } \right)& = &\sum\limits_{n = 1}^{N_S \left( {{}^1A_{1g} } \right)} {{{\left| {t_{}^{0,ns} \left( {R_{ij} } \right)} \right|^2 } \mathord{\left/
 {\vphantom {{\left| {t_{}^{0,ns} \left( {R_{ij} } \right)} \right|^2 } {\Delta _{nS} }}} \right.
 \kern-\nulldelimiterspace} {\Delta _{nS} }}}  - \label{eq:8} \\
&-&\sum\limits_{m = 1}^{N_T \left( {{}^3A_{1g} } \right)} {{{\left| {t_{}^{0,mT} \left( {R_{ij} } \right)} \right|^2 } \mathord{\left/
 {\vphantom {{\left| {t_{}^{0,mT} \left( {R_{ij} } \right)} \right|^2 } {\Delta _{mT} }}} \right.
 \kern-\nulldelimiterspace} {\Delta _{mT} }}} ,\;\;
\nonumber
\end{eqnarray}

\begin{eqnarray}
\Delta J_{FM}^{B_{1g} } \left( {\vec R_{ij} } \right) &=& \sum\limits_{n = 1}^{N_S \left( {{}^1B_{1g} } \right)} {{{\left| {t_{}^{0,ns} \left( {\vec R_{ij} } \right)} \right|^2 } \mathord{\left/
 {\vphantom {{\left| {t_{}^{0,ns} \left( {\vec R_{ij} } \right)} \right|^2 } {\Delta _{nS} }}} \right.
 \kern-\nulldelimiterspace} {\Delta _{nS} }}}  - \nonumber \\
&-&\sum\limits_{m = 1}^{N_T \left( {{}^3B_{1g} } \right)} {{{\left| {t_{}^{0,mT} \left( {\vec R_{ij} } \right)} \right|^2 } \mathord{\left/
 {\vphantom {{\left| {t_{}^{0,mT} \left( {\vec R_{ij} } \right)} \right|^2 } {\Delta _{mT} }}} \right.
 \kern-\nulldelimiterspace} {\Delta _{mT} }}} ,\;
\nonumber
\end{eqnarray}

\begin{figure}
\includegraphics{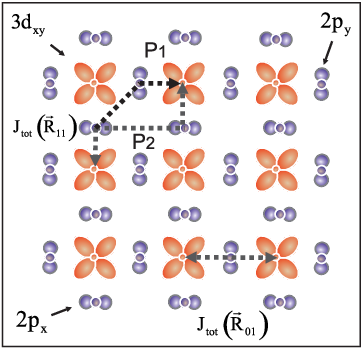}
\caption{Paths $P_0$, $P_1$   and $P_2$   of the superexchange interactions $J_{tot} \left( {\vec R_{01} } \right)$ and $J_{tot} \left( {\vec R_{11} } \right)$ for the nearest and next-nearest neighbors. Here, the oxygen $2p$ orbitals $\pi$-overlap with the $t_{2g}$ magnetic ions, and form $90^\circ \left( {P_1 } \right)$ and  $\sigma \left( {P_2 } \right)$ - overlapping between themselves. The  interactions $J_{tot}^{\left( {P_1 } \right)} \left( {\vec R_{11} } \right)$ and  $J_{tot}^{\left( {P_2 } \right)} \left( {\vec R_{11} } \right)$  are comparable in magnitude for magnetic materials with the partially occupied  $t_{2g} $
shell. }
\label{fig:4}
\end{figure}

The contributions $\Delta J_{AFM}^{A_{1g} } \left( {\vec R_{ij} } \right)$ and $\Delta J_{FM}^{B_{1g} } \left( {\vec R_{ij} } \right)$ have AFM and FM nature, respectively, due to the levels of the two-hole spin triplets $\left| {{}^3A_{1g} } \right\rangle $ and singlets $\left| {{}^1B_{1g} } \right\rangle $ lying higher in energy than the levels of the spin singlets and triplets, respectively. In fact, from Eq.(\ref{eq:8}), one finds
\begin{eqnarray}
J_{tot} \left( {\vec R_{01} } \right) &=& \Delta J_{AFM}^{A_{1g} } \left( {\vec R_{01} } \right) + \Delta J_{FM}^{B_{1g} } \left( {\vec R_{01} } \right) \approx \label{eq:9} \\
&\approx& \left(10.4-0.5\right)\times10^{-2}~eV=9.9\times10^{-2}eV \nonumber \\
J_{tot} \left( {\vec R_{11} } \right) &=& \Delta J_{AFM}^{A_{1g} } \left( {\vec R_{11} } \right) + 0 \approx 0.2\times10^{-2}eV,
\nonumber
\end{eqnarray}
where $\Delta J_{AFM}^{A_{1g} } \left( {\vec R_{01} }\right)\approx\left(15.6-5.2\right)\times10^{-2}=10.4\times10^{-2}eV$, $\Delta J_{FM}^{B_{1g} } \left( {\vec R_{01} } \right)\approx \left(0.4-0.9\right)\times10^{-2}eV=-0.5eV$ and $\Delta J_{FM}^{B_{1g} } \left( {\vec R_{11} } \right) = 0$, in the square CuO$_2$ layer, due to the zero diagonal mobility of the virtual holes in any of the two-hole state with the  ${B_{1g} }$ point symmetry. Consequently, there is only the nonzero AFM contribution to the diagonal superexchange $J_{tot} \left( {\vec R_{11} } \right)$ in Fig.\ref{fig:3}, due to the invariance of the square CuO$_2$ lattice upon the replacement $x\leftrightarrows y$.

\section{\label{sec:IV} Discussion and conclusions\\}

Magnitudes of the diagonal AFM superexchange interaction in the Eq.(\ref{eq:9}) qualitatively coincide with the results of the work~\cite{Annet_etal1989} based on the generalized Hubbard hamiltonian, but contradict the FM interaction~\cite{Wan2009} in the full potential linearized-muffin-tin-orbital  method, the density-functional theory DFT, where the diagonal AFM interaction was obtained only for the doped cuprates. Unlike the authors of the work~\cite{Wan2009}, we do not think that the differences between the results of the works~\cite{Annet_etal1989} and ~\cite{Wan2009} are related to the choice of method.  In the work~\cite{Wan2009}, real structures of cuprates were investigated, where there is indeed a non-zero FM contribution. Unfortunately, the authors do not discuss a change in the nature of the diagonal exchange from FM to AFM at doping level $x = 0.3$.  This effect can also be related to the structural sensitivity of the FM contribution. The calculation~\cite{Wan2009} of the exchange interaction in parent and doped cuprates with subsequent study of the dependence of $T_C$ on $J$ within the $t-J$ model is not self-consistent. Indeed, the $t-J$ model was derived from the single-band Hubbard model with the two-particle spin singlets~\cite{Spalek2007}(two-hole Zhang-Rice singlets in the CuO$_2$ layer~\cite{Jefferson1990}), therefore values of $J$ in the $t-J$ model can only correspond to the AFM interaction. The non-zero FM contribution has been previously studied in more complex models~\cite{Chao1977, Spalek1980} with two-particle spin triplets. The FM contribution  in the square CuO$_2$ layer of doped cuprates is possible only when taking into account the double exchange and RKKI interactions. However, as the authors~\cite{Wan2009} note, at the current doping levels, the contribution from the latter is small. Probably, the conclusion about the absence of correlation between $J$ and $T_C$ in a framework of the non-self-consistent approach, where the exchange constant $J$ and its effects on $T_C$ are calculated in different DFT and $t-J$ model approaches, requires additional study. The models  $t - t' - J$, $t - t' - t'' - J$ with extended hopping also based on the single-band approach do not contain any FM contributions, and therefore, they correctly describe magnetic interactions only in a hypothetical square lattice rather than in the real CuO$_2$ layer with the broken square symmetry (e.g. with tilted CuO$_6$ octahedra in the D and U stripes~\cite{Bianconi_etal1996}) which are controlled by spatially heterogeneous lattice microstrain ~\cite{Bianconi_2000, Albertini_2023}. The internal chemical pressure in doped perovskites gives nanoscale phase separation ~\cite{Kugel_2008} and superlattices ~\cite{Sboychakov_2022}. In fact, strain uncovers the interplay between two- and three-dimensional charge densities ~\cite{Vinograd_2024}, and between the lattice superstructures and the electronic structure of cuprate perovskites ~\cite{Hameed_2024}.

The FM interaction with the second neighbors indeed better matches the experimental spin-wave dispersions for La$_2$CuO$_4$~\cite{Wan2009}. As follows from our results, this can be observed only in a real cuprate material with broken square symmetry. We do not expect any peculiarities in the spin wave spectrum that could be calculated in Green's function approach with the Hubbard operators and Tyablikov decoupling~\cite{Valkov1982}.  The largest difference between our results and ~\cite{Wan2009} will be observed for small $S=1/2$, due to the Goldstein-Primakov representation using the $1/S$ expansion. However, there are relevant aspects to spin wave  studies where we can make some corrections. The superexchange interaction is a superposition of contributions from all possible virtual electron-hole pairs in the Eq.(\ref{eq:8}). Each of  contributions has its own small parameter $t/U$. Therefore, it is not surprising that fitting the magnon dispersion using the linear spin-wave theory leads to systematic errors in the estimates of the exchange parameters and corresponding overestimations of $t/U$, which has a slightly different physical meaning. Along with the observed magnon dispersion on magnetic Brillouin zone boundary, such discrepancies in the $t/U$ parameter were one of reasons for introducing four-spin cycling exchange into the physics of cuprates~\cite{Bao_2025}. We also confirm that there are valid reasons for considering magnetic frustrations~\cite{Annet_etal1989} in a square lattice. One can expect that in most cases the frustrations will be negligible due to the small ratio ${{J_{tot} \left( {\vec R_{11} } \right)} \mathord{\left/
 {\vphantom {{J_{tot} \left( {\vec R_{11} } \right)} {J_{tot} \left( {\vec R_{01} } \right)}}} \right.
 \kern-\nulldelimiterspace} {J_{tot} \left( {\vec R_{01} } \right)}} \approx ~0.016$ ($0.08$  in the work~\cite{Annet_etal1989}). Therefore, the transfer of the results obtained for the pseudogap at the Mott transition in the triangular lattice Hubbard model with next-nearest-neighbor hopping and magnetic frustrations~\cite{Downey2023} to the square CuO$_2$ layer is just motivating.

 To summarize, in this paper we have obtained  $J_{tot}^{} \left( {\vec R_{01} } \right) \approx 9.9\times10^{-2}~eV$ with the parameters of hamiltonian used earlier to calculate the energy structure and angle resolved photoemission spectra of cuprates~\cite{Korshunov_etal2005}. The diagonal $J_{tot} \left( {\vec R_{11} } \right)\approx 0.2\times10^{-2}eV$ superexchange interaction in the simple square lattice of the CuO$_2$ layer always has the AFM nature due to the  symmetry prohibition on the FM contribution $\Delta J_{FM}^{B_{1g} } \left( {\vec R_{11}} \right) = 0$. However, there is no prohibition $\Delta J_{FM}^{B_{1g} }\left( {\vec R_{01}} \right) \approx -0.5\times 10^{-2}~eV$ and $\Delta J_{FM}^{A_{1g} }\left( {\vec R_{01}} \right) \approx 10.4\times10^{-2}~eV$ for the interacting nearest neighbors. Our calculation is limited to five orbitals, but the all possible electron-hole pairs contribute to the superexchange $\hat{H}_S$ interaction. Beyond the orbital limit, actual $4s$ orbitals~\cite{Pavarini_etal2001} can lead to a contribution only to the $\Delta J_{FM}^{B _{1g} } \left( {\vec R_{ij}} \right)$ interaction in the Fig.\ref{fig:3}, i.e. to the increasing FM contribution. Let us also note that a type of magnetic ions remains clearly important in relation to the prohibition.  This can be seen in the Fig.\ref{fig:4}, where for magnetic ions with a partially occupied  $t_{2g}$  shell, the overlapping $2p$ orbitals of oxygen ions along path P$_2$ is quite significant and should be taken into account in calculating the diagonal superexchange constant  $J_{tot} \left( {\vec R_{11} } \right)$.

Further, it would be of interest to consider the effect of a certain type of
broken square $x\leftrightarrows y$ symmetry with unequal lattice parameters between the orthorhombic $a$ and $b$ axes on the experimentally observed "Y shift" with a surprisingly large tilting angle (the so called diagonal stripes) ~\cite{Wei2024, Bianconi_etal1996, Bianconi_2000, Albertini_2023, Kugel_2008, Sboychakov_2022}.
\begin{acknowledgments}
The work on sections 1-4 of the paper was carried out  with the support of the RSF grant No. 24-12-00044. The appendices A and B were carried out within a framework of
the scientific topic of the state assignment of L.V. Kirensky Institute of Physics, SB RAS.

\end{acknowledgments} $\\$

\appendix

\section{The symmetric cell representation for the $pd$ model}
To obtain the energy spectrum shown in Fig.\ref{fig:2}a we start from the multiorbital $pd$ Hamiltonian:
$\hat H = {\hat H_d} + {\hat H_p} + {\hat H_{pd}} + {\hat H_{pp}}$, where
\begin{widetext}
\begin{eqnarray}
{\hat {H_d}} = \sum\limits_{i\lambda \sigma } {\left[ {\left( {{\varepsilon _\lambda } - \mu } \right)d_{\lambda i\sigma }^ + {d_{\lambda i\sigma }} + \frac{{{U_d}}}{2}\hat n_{\lambda i}^\sigma \hat n_{\lambda i}^{ - \sigma } + \frac{1}{2}\sum\limits_{\lambda ' \ne \lambda } {\left( {\sum\limits_{\sigma '} {{V_{\lambda \lambda '}}\hat n_{\lambda i}^\sigma \hat n_{\lambda 'i}^{\sigma '}}  - {J_H}d_{\lambda i\sigma }^ + {d_{\lambda i\bar \sigma }}d_{\lambda 'i\bar \sigma }^ + {d_{\lambda 'i\sigma }}} \right)} } \right]},
\label{eq:A.1} \\
{\hat H_p} = \sum\limits_{m\alpha \sigma } {\left[ {\left( {{\varepsilon _\alpha } - \mu } \right)p_{\alpha m\sigma }^ + {p_{\alpha m\sigma }} + \frac{{{U_p}}}{2}\hat n_{\alpha m}^\sigma \hat n_{\alpha m}^{ - \sigma } + \frac{1}{2}\sum\limits_{\alpha ' \ne \alpha ,\sigma '} {{V_{\alpha \alpha '}}\hat n_{\alpha m}^\sigma \hat n_{\alpha 'm}^{\sigma '}} } \right]}, \nonumber \\
{\hat H_{pd}} = \sum\limits_{mi} {\sum\limits_{\alpha \lambda \sigma } {\left[ {t_{im}^{\lambda \alpha }\left( {p_{\alpha m\sigma }^ + {d_{\lambda f\sigma }} + h.c.} \right) + \frac{{V_{im}^{pd}}}{2}\sum\limits_{\sigma '} {\hat n_{\alpha m}^\sigma \hat n_{\lambda i}^{\sigma '}} } \right]} }, ~
{\hat H_{pp}} = \sum\limits_{mn} {\sum\limits_{\alpha \beta \sigma } {t_{mn}^{\alpha \beta }\left( {p_{\alpha m\sigma }^ + {p_{\beta n\sigma }} + h.c.} \right)} }.\nonumber
\end{eqnarray}
\end{widetext}
Here, $n_{\lambda i}^\sigma  = d_{\lambda i\sigma }^ + {d_{\lambda i\sigma }}$, $n_{\alpha m}^\sigma  = p_{\alpha m\sigma }^ + {p_{\alpha m\sigma }}$, where the indices $i(j)$  and  $m(n)$ run over all positions ${d_\lambda } = {d_{{x^2} - {y^2}}},{d_{3{r^2} - {z^2}}}$  and  ${p_\alpha } = {p_x},{p_y},{p_z}$(- apical) localized one electron states with energies ${\varepsilon _\lambda }$  and ${\varepsilon _\alpha}$; ${t^{\lambda \alpha }_{im}}$   and ${t^{\alpha \beta }_{mn}}$  the hopping matrix elements; ${U_d }$, ${U_p }$    and  ${J_H}$ are one site Coulomb interactions and the Hund exchange interaction, ${V^{pd}_{im}}$  is the energy of repulsion of cation and anion electrons. A correct transition from the $pd$-Hamiltonian (\ref{eq:A.1})  to the Eq.(\ref{eq:A.12}) in the multielectron  representation of the Hubbard operators~\cite{Hubbard_1963} is possible when constructing well localized Wannier cell oxygen states $\left| {p_{\lambda i\sigma }^ + } \right\rangle$ in the Eq.(\ref{eq:5}). Note, here and below a prime denotes interactions involving apical oxygen  ions. After this step the $pd$-Hamiltonian becomes a sum of
intracell and intercell terms ~\cite{Shastry_1989, Feiner_etal1996, Gavrichkov_2000}:

\begin{widetext}
\begin{eqnarray}
&&\hat H =\hat H_c  + \hat H_{cc}, \hat H_c  = \sum\limits_{i\sigma } {\hat H_{i\sigma } },
~\hat H_{i\sigma }  = \hat h_{i}^{\left( b \right)}  + \hat h_{i}^{\left( a \right)}  + \hat h_{i}^{\left( {ab} \right)} \label{eq:A.2} \\
&&\hat h_i^{\left( b \right)}  = \left( {\varepsilon _b n_b^\sigma   + \varepsilon _{d_x } n_{d_x }^\sigma  } \right) + {\textstyle{1 \over 2}}U_d n_{d_x }^\sigma  n_{d_x }^{ - \sigma }  + {\textstyle{1 \over 2}}U_b n_b^\sigma  n_b^{ - \sigma }  + \sum\limits_{\sigma '} {V_{pd} n_{d_x }^\sigma  n_b^{\sigma '} }  - \tau _b \sum\limits_\sigma  {\left( {d_{x\sigma }^ +  b_\sigma   + h.c.} \right)} \nonumber \\
&&\hat h_i^{\left( a \right)}  = \left( {\varepsilon _a n_a^\sigma   + \varepsilon _{d_z } n_{d_z }^\sigma   + \varepsilon _{p_z } n_{p_z }^\sigma  } \right) + {\textstyle{1 \over 2}}U_d n_{d_z }^\sigma  n_{d_z }^{ - \sigma }  + {\textstyle{1 \over 2}}U_a n_a^\sigma  n_a^{ - \sigma }  + {\textstyle{1 \over 2}}U'_p n_{p_z }^\sigma  n_{p_z }^{ - \sigma }  + \sum\limits_{\sigma '} {\left( {V'_{pd} n_{d_z }^\sigma  n_{p_z }^{\sigma '}  + V_{pd} n_{d_z }^\sigma  n_a^{\sigma '} } \right) + } \nonumber \\
&&+ \tau _a \left( {d_{z\sigma }^ +  a_\sigma   + h.c} \right) - \tau '_{pd} \left( {d_{z\sigma }^ +  p_{z\sigma }  + h.c.} \right) - t'_{pp} \left( {a_\sigma ^ +  p_{z\sigma }  + h.c} \right) \nonumber \\
&&\hat h_i^{\left( {ab} \right)}  = \sum\limits_{\sigma '} {U_d n_{d_x }^\sigma  n_{d_z }^{\sigma '}  + U_{ab} n_a^\sigma  n_b^{\sigma '}  + V_{pd} n_{d_x }^\sigma  n_a^{\sigma '}  + V_{pd}^{} n_b^\sigma  n_{d_z }^{\sigma '}  + V'_{pd} n_{d_x }^\sigma  n_{p_z }^{\sigma '} } \nonumber \\
&&\hat H_{cc}  = \sum\limits_{{i \ne j}} {\sum\limits_\sigma  {\left( {\hat h_{ij}^{\left( b \right)}  + \hat h_{ij}^{\left( a \right)}  + \hat h_{ij}^{\left( {ab} \right)} } \right)} } \nonumber \\
&&\hat h_{ij}^{\left( b \right)}  =  - 2t_{pd} \mu _{ij} \left( {d_{xi\sigma }^ +  b_{j\sigma }  + b_{i\sigma }^ +  d_{xi\sigma } } \right) - 2t_{pp} \nu _{ij} b_{i\sigma }^ +  b_{j\sigma } \nonumber \\
&&\hat h_{ij}^{\left( a \right)}  = \frac{{2t_{pd} }}{{\sqrt 3 }}\lambda _{ij} \left( {d_{zi\sigma }^ +  a_{j\sigma }  + h.c.} \right) + 2t_{pp} \nu _{ij} a_{i\sigma }^ +  a_{j\sigma }  - 2t'_{pp} \lambda _{ij} \left( {p_{zi\sigma }^ +  a_{j\sigma }  + h.c.} \right) \nonumber \\
&&\hat h_{ij}^{\left( {ab} \right)}  = \frac{{2t_{pd} }}{{\sqrt 3 }}\xi _{ij} \left( {d_{zi\sigma }^ +  b_{j\sigma }  + h.c.} \right) + 2t_{pp} \chi _{ij} \left( {a_{i\sigma }^ +  b_{j\sigma }  + h.c.} \right) - 2t'_{pp} \xi _{ij} \left( {p_{zi\sigma }^ +  b_{j\sigma }  + h.c.} \right), \nonumber
\end{eqnarray}
\end{widetext}
where $\varepsilon _b  = \varepsilon _p  - 2t_{pp} \nu _{00}$, $\varepsilon _a  = \varepsilon _p  + 2t_{pp} \nu _{00}$, $\tau _b  = 2t_{pd} \mu _{00}$, $\tau _a  = 2t_{pd} \lambda _{00} /\sqrt 3$ and $\tau'_{pd}  = 2t'_{pd} /\sqrt 3$, $\tau'_{pp}  = 2t'_{pp} \lambda _{00}$.

As the next step, we shall obtain the eigenvalues
and eigenstates of the single-cell Hamiltonian $\hat{H}_c$. In the vacuum
sector $N_+(d^{10})$ we have the proper state $d^{10}p^6$ or $\left| 0 \right\rangle$. In the
single-hole $b_{1g}$ sector on the basis $\left|  {d_{xs }} \right\rangle$ and
$\left| {b^s_{1g}} \right\rangle$ states the eigenvectors
$\left| {b^s_{1g}} \right\rangle_q  = \beta _q \left( b\right)\left| {b_s} \right\rangle  + \beta _q \left( {d_x } \right)\left| {d_{xs}} \right\rangle$ with energies $\varepsilon _{b_{1g},q }$ can be found by exact diagonalization of $\hat h_i^{\left( b \right)}$:

\begin{equation}
\hat h_i^{\left( b \right)}  = \left( {\begin{array}{*{20}c}
   {\varepsilon _{d_x } } & { - \tau _b }  \\
   { - \tau _b } & {\varepsilon _b }  \\
\end{array}} \right)
\label{eq:A.3}
\end{equation}
In the single hole $a_{1g}$ sector in the basis $\left| a_s\right\rangle$, $\left| p_{zs}\right\rangle$,
and $\left| d_{zs} \right\rangle$ states, the eigenvectors $\left| {a^s_{1g} } \right\rangle_q  = \alpha _q \left( a \right)\left| a_s   \right\rangle  + \alpha _q \left( {p_z } \right)\left| p_{zs}\right\rangle  + \alpha _q \left( {d_z } \right)\left| d_{zs}^ +   \right\rangle$ with energies $\varepsilon _{a_{1g},q }$ can be found by exact diagonalization of $\hat h_i^{\left( a \right)}$

\begin{equation}
\hat h^{\left( a \right)}  = \left( {\begin{array}{*{20}c}
   {\varepsilon _{d_z } } & {\tau _a } & { - \tau '_{pd} }  \\
   {\tau _a } & {\varepsilon _{_a } } & { - t'_{pp} }  \\
   { - \tau '_{pd} } & { - t'_{pp} } & {\varepsilon _{p_z } }  \\
\end{array}} \right)
\label{eq:A.4}
\end{equation}
The eigenstates of a cell in the two-hole $A_{1g}$ sector
$\left| A_{1g} \right\rangle_{nS}  = \sum\limits_p {A_{nS,p} \left| {A_p } \right\rangle }$, where the coefficients are the eigenvectors
$A_{nS,p}$, and the set of the basis singlet $\left| {A_p } \right\rangle$
states  are presented in the Tab.\ref{tab:A.2}. The eigenstates $\left| A_{1g}  \right\rangle_{nS}$ with energy $\varepsilon _{nS}$ can be found by exact diagonalization
of the matrix $\hat h_i^{\left( A_{1g} \right)}$

\begin{equation}
\hat h^{\left( A_{1g} \right)}  = \left( {\begin{array}{*{20}c}
   {\hat h_{11}^{\left( A \right)} } & 0  \\
   0 & {\hat h_{22}^{\left( A \right)} }  \\
\end{array}} \right)
\label{eq:A.5}
\end{equation}
where
\begin{equation}
\hat h^{\left( A_{1g} \right)} _{11}  = \left( {\begin{array}{*{20}c}
   {\varepsilon _b  + \varepsilon _{d_x }  + V_{pd} } & { - \sqrt 2 \tau _b } & {\sqrt 2 \tau _b }  \\
   { - \sqrt 2 \tau _b } & {2\varepsilon _b  + U_b } & 0  \\
   { - \sqrt 2 \tau _b } & 0 & {2\varepsilon _{d_x }  + U_d }  \\
\end{array}} \right)
\label{eq:A.6}
\end{equation}
and
\begin{widetext}
\begin{equation}
\hat h_{22}^{\left( A_{1g} \right)}  = \left( {\begin{array}{*{20}c}
   {\varepsilon _a  + \varepsilon _{p_z }  + V'_p } & { - \tau '_{pd} } & {\tau _a } & { - \sqrt 2 t'_{pp} } & { - \sqrt 2 t'_{pp} } & 0  \\
   { - \tau '_{pd} } & {\varepsilon _{d_z }  + \varepsilon _a  + V_{pd} } & { - t'_{pp} } & {\sqrt[{}]{2}\tau _a } & 0 & {\sqrt[{}]{2}\tau _a }  \\
   {\tau _a } & { - t'_{pp} } & {\varepsilon _{d_z }  + \varepsilon _{p_z }  + V'_{pd} } & 0 & { - \sqrt 2 \tau '_{pd} } & { - \sqrt 2 \tau '_{pd} }  \\
   { - \sqrt 2 t'_{pp} } & {\sqrt[{}]{2}\tau _a } & 0 & {2\varepsilon _a  + U_a } & 0 & 0  \\
   { - \sqrt 2 t'_{pp} } & 0 & { - \sqrt 2 \tau '_{pd} } & 0 & {2\varepsilon _{p_z }  + U'_p } & 0  \\
   0 & {\sqrt[{}]{2}\tau _a } & { - \sqrt 2 \tau '_{pd} } & 0 & 0 & {2\varepsilon _{d_z }  + U_d }  \\
\end{array}} \right)
\label{eq:A.7}
\end{equation}
\end{widetext}

\begin{table}
\caption{Eigenvectors $A_{nS,p}$ and the set of basis singlet functions
$\left| {A_p} \right\rangle$ \label{tab:A.2}}
\begin{ruledtabular}
\begin{tabular}{@{}lc@{}}
$A_{nS,p}$ & $\left| {A_p } \right\rangle $ \\
\hline \\
\footnotesize    $A_{nS,1}(d_xb)$   &\footnotesize     $\left| {ZR} \right\rangle  = {\textstyle{1 \over {\sqrt 2 }}}\left| {d_{x \downarrow }^ +  b_ \uparrow ^ +   - d_{x \uparrow }^ +  b_ \downarrow ^ +  } \right\rangle$ \\
$A_{nS,2}(bb)$ & $\left| {b_ \downarrow ^ +  b_ \uparrow ^ +} \right\rangle $ \\
$A_{nS,3}(d_x,d_x)$ & $\left| {d_{x \downarrow }^ +  d_{x \uparrow }^ + }\right\rangle $ \\
$A_{nS,4}(p_z,a)$ & ${\textstyle{1 \over {\sqrt 2 }}}\left| {p_{z \downarrow }^ +  a_ \uparrow ^ +   - p_{z \uparrow }^ +  a_ \downarrow ^ + } \right\rangle$ \\
$A_{nS,5}(d_z,a)$ & ${\textstyle{1 \over {\sqrt 2 }}}\left| {d_{z \downarrow }^ +  a_ \uparrow ^ +   - d_{z \uparrow }^ +  a_ \downarrow ^ + } \right\rangle$ \\
$A_{nS,6}(d_z,p_z)$ & ${\textstyle{1 \over {\sqrt 2 }}}\left| {d_{z \downarrow }^ +  p_{z \uparrow} ^ +   - d_{z \uparrow }^ +  p_{z \downarrow} ^ +  } \right\rangle$ \\
$A_{nS,7}(aa)$ & $\left| {a_ \downarrow ^ +  a_ \uparrow ^ +} \right\rangle $ \\
$A_{nS,8}(p_zp_z)$ & $\left| {p_{z \downarrow} ^ +  p_{z \uparrow} ^ +} \right\rangle $ \\
$A_{nS,9}(d_zd_z)$ & $\left| {d_{z \downarrow} ^ +  d_{z \uparrow} ^ +} \right\rangle $ \\
\end{tabular}
\end{ruledtabular}
\end{table}

In the two-hole triplet sector $^3B_{1g}$ we obtain the
eigenvectors $\left| {B^T_{1g} } \right\rangle_{m}  = \sum\limits_p {B_{mT,p} \left| {B_{p} } \right\rangle }$, where the corresponding coefficients $B_{mT,p}$
and the set of basis functions $\left| {B_{p} } \right\rangle$ are presented in the Tab.\ref{tab:A.3}
with energies $\varepsilon_{mT}$ found by diagonalizing the matrix $\hat h_{}^{\left( {^3B_{1g}} \right)}:$
\begin{table*}
\caption{Eigenvectors $B_{mT,p}$ and the set of basis states $
\left| {B_{pT} } \right\rangle$. \label{tab:A.3}}
\begin{ruledtabular}
\begin{tabular}{@{}lccc@{}}
$B_{mT,p}$ & $\left| {B_{pT=-1} } \right\rangle $  & $\left| {B_{pT=0} } \right\rangle $ & $\left| {B_{pT=+1} } \right\rangle $ \\
\hline \\
\footnotesize    $B_{mT,1}(d_xa)$   & \footnotesize     $
\left| {d_{x \downarrow }^ +  a_ \downarrow ^ +  } \right\rangle$ & ${\textstyle{1 \over {\sqrt 2 }}}\left| {d_{x \downarrow }^ +  a_ \uparrow ^ +   + d_{x \uparrow }^ +  a_ \downarrow ^ +  } \right\rangle$ & $\left| {d_{x \uparrow }^ +  a_ \uparrow ^ +  } \right\rangle$ \\
$B_{mT,2}(ba)$ & $\left| {b_ \downarrow ^ +  a_ \downarrow ^ +  } \right\rangle$ &  ${\textstyle{1 \over {\sqrt 2 }}}\left| {b_ \downarrow ^ +  a_ \uparrow ^ +   + b_ \uparrow ^ +  a_ \downarrow ^ +  } \right\rangle$ & $\left| {b_ \uparrow ^ +  a_ \uparrow ^ + } \right\rangle$ \\
$B_{mT,3}(d_x,d_z)$ & $\left| {d_{x \downarrow }^ +  d_{z \downarrow }^ +  } \right\rangle$ & ${\textstyle{1 \over {\sqrt 2 }}}\left| {d_{x \downarrow }^ +  d_{z \uparrow }^ +   + d_{x \uparrow }^ +  d_{z \downarrow }^ +  } \right\rangle$ & $\left| {d_{x \uparrow }^ +  d_{z \uparrow }^ +  } \right\rangle$ \\
$B_{mT,4}(d_z,b)$ & $\left| {d_{z \downarrow }^ +  b_ \downarrow ^ +  } \right\rangle$ &  ${\textstyle{1 \over {\sqrt 2 }}}\left| {d_{x \downarrow }^ +  b_ \uparrow ^ +   + d_{x \uparrow }^ +  b_ \downarrow ^ +  } \right\rangle$ & $\left| {d_{z \uparrow }^ +  b_ \uparrow ^ +  } \right\rangle$ \\
$B_{mT,5}(d_x,p_z)$ & $\left| {d_{x \downarrow }^ +  p_{z \downarrow }^ +  } \right\rangle$ & ${\textstyle{1 \over {\sqrt 2 }}}\left| {d_{x \downarrow }^ +  p_{z \uparrow }^ +   + d_{x \uparrow }^ +  p_{z \downarrow }^ +  } \right\rangle$ & $\left| {d_{x \uparrow }^ +  p_{z \uparrow }^ + } \right\rangle$ \\
$B_{mT,6}(b,p_z)$ & $\left| {b_ \downarrow ^ +  p_{z \downarrow }^ + } \right\rangle$ & ${\textstyle{1 \over {\sqrt 2 }}}\left| {b_ \downarrow ^ +  p_{z \uparrow }^ +   + b_ \uparrow ^ +  p_{z \downarrow }^ +  } \right\rangle$ & $\left| {b_ \uparrow ^ +  p_{z \uparrow }^ + } \right\rangle$ \\
\end{tabular}
\end{ruledtabular}
\end{table*}

\begin{widetext}
\begin{equation}
\hat h_{i}^{\left( {^3B_{1g}} \right)}  = \left( {\begin{array}{*{20}c}
   {\varepsilon _a  + \varepsilon _{d_x }  + V_{pd} } & { - \tau _b } & {\tau _a } & 0 & { - t'_{pp} } & 0  \\
   { - \tau _b } & {\varepsilon _a  + \varepsilon _b  + U_b } & 0 & {\tau _a } & 0 & { - t'_{pp} }  \\
   {\tau _a } & 0 & {\varepsilon _{d_z }  + \varepsilon _{d_x }  + U_d } & { - \tau _b } & { - \tau '_{pd} } & 0  \\
   0 & {\tau _a } & { - \tau _b } & {\varepsilon _{d_z }  + \varepsilon _b  + V_{pd} } & 0 & { - \tau'_{pd} }  \\
   { - t'_{pp} } & 0 & { - \tau'_{pd} } & 0 & {\varepsilon _{d_x }   + \varepsilon _{p_z }  + V'_{pd} } & { - \tau _b }  \\
   0 & { - t'_{pp} } & 0 & { - \tau'_{pd} } & { - \tau _b } & {\varepsilon _b  + \varepsilon _{p_z }  + V'_{pp} }  \\
\end{array}} \right)
\label{eq:A.8}
\end{equation}
\end{widetext}

A diagonalization of  the intracell path $\hat H_c$ for a CuO$_6$ cluster
is done separately in the different configuration sectors: $N_+$, $N_0$, $N_-$ with $0$, $1$, and $2$ holes per cell. The vacuum section $N_+$ corresponds to the $p^6d^{10}$ configuration. The matrices ~ $\hat h_{i}^{\left( {^3A_{1g}} \right)}$ and ~ $\hat h_{i}^{\left( {^1B_{1g}} \right)}$ are diagonalized in the same way as above. In the new basis any single-electron $\rho_{i\lambda \sigma }$ operators become
\begin{eqnarray}
&\rho^{(+)}_{i\lambda \sigma }& = \sum\limits_r {\gamma _{\lambda \sigma }^{} \left( r \right)X_{i}^{(+)r} }= \label{eq:A.9}\\
&=&\gamma _{\lambda \sigma } (^1A,b^{\sigma}_{1g})X_i^{^1A,b^{\sigma}_{1g} }  + \sum\limits_n {\gamma _{\lambda \sigma } (b^{- \sigma}_{1g},\;nS)X_i^{ b^{- \sigma}_{1g},\;nS} }  + \nonumber \\ &&+\sum\limits_m {\gamma _{\lambda \sigma } (b^{\sigma}_{1g},mT)} \left( {X_i^{b^{\sigma}_{1g},\;m2\sigma }  + \frac{1}{{\sqrt 2 }}X_i^{ b^{-\sigma}_{1g},\;m0} } \right),
\nonumber
\end{eqnarray}

where the $n$ and $m$$(T=0,2\sigma)$ indices run over all two-hole spin singlet and triplet states, and $\rho_{\lambda i\sigma }  = d_{xi\sigma } ,d_{zi\sigma } ,a_{i\sigma } ,b_{i\sigma } ,p_{zi\sigma }$ and $r$ is the index
of root vector $(qq')$. Here, to make it easier to
work with Hubbard operators~\cite{Hubbard_1963}, we employ Zaitsev's notation
~\cite{Zaitsev_1976}, where to each pair (initial and final) of
states $\left| q \right\rangle  \to \left| q' \right\rangle$
there is associated a root vector $r$, so that
\begin{equation}
X_{i}^{{qq'} }  \to X_{i}^r
\label{eq:A.10}
\end{equation}

The matrix elements of the hopping amplitudes $\gamma _{\lambda \sigma } \left( r \right)$
corresponding to these root vectors,
can be calculated directly by performing an exact diagonalization
of the intracell path $H_c$ and are presented here.

\begin{eqnarray}
&&\gamma _{\lambda \sigma }(b_{1g \sigma'},^1A_{1g,nS})=\eta \left( \sigma  \right)\left( {\delta _{\sigma \sigma '}-1} \right)\sum\limits_{\lambda'=b,d_x} {\beta \left( {\lambda'} \right)A_{nS} \left( {\lambda \lambda'} \right)} \nonumber \\
&&\gamma _{\lambda \sigma }(b_{1g \sigma'},^3B_{1g,mT=\pm1})=-\delta _{\sigma \sigma '} \sum\limits_{\lambda'=d_x,b} {\beta \left( {\lambda'} \right)} B_{mT} \left( {d_x \lambda' } \right) \nonumber \\
&&\gamma _{\lambda \sigma }(b_{1g \sigma'},^3B_{1g,mT=0})={\textstyle{1 \over {\sqrt 2 }}}\left( {\delta _{\sigma \sigma '} }-1
\right) \sum\limits_{\lambda'=d_x,b} {\beta \left( {\lambda'} \right)} B_{mT} \left( {\lambda \lambda' } \right) \nonumber \\
\label{eq:A.11}
\end{eqnarray}
with $\lambda=d_x, b$ in first line and $\lambda=d_z, a, p_z$ in the second and third lines respectively.
Only the bottom state $|b_{1g}\rangle$ is taken into account in the single-hole sector $N_0(d^9)$ and all states $|^1A_{1g}\rangle_{nS}$, $|^3B_{1g}\rangle_{mT}$, $|^1B_{1g}\rangle_{nS}$, $|^3A_{1g}\rangle_{mT}$ are taken into account in the two-hole $N_-(d^8)$ sector.
As a result the intracell $H_c$ and intercell $H_{cc}$ paths of Hamiltonian in the representation of the Hubbard operators take the form of the Eq.(\ref{eq:A.12})~\cite{ Ovchinnikov_etal2004, Ovchinnikov_etal2012}

\begin{widetext}
\begin{eqnarray}
&\hat H_c  &= \sum\limits_i {\left\{ {\left( {E_{{}^1A}  - N_ +  \mu } \right)X_i^{{}^1A{}^1A}  + \left( {\varepsilon _{b_{1g} }  - N_0 \mu } \right)\sum\limits_{s}{X_i^{b_{1g}^{s} b_{1g}^{s} }}  + \sum\limits_{h = nS,mT}^{} {\left( {E_h  - N_\_ \mu } \right)X_i^{hh} } } \right\}} \label{eq:A.12} \\
&\hat H_{cc}  &= \sum\limits_{ij\sigma } {\sum\limits_{rr'} {t_\sigma ^{rr'} \left( {\vec R_{ij} } \right)} X_i^{+r}  } X_j^{r'} \nonumber \\
&t_\sigma ^{r,r'}& \left( {\vec R_{ij} } \right) = \sum\limits_{ij} {\sum\limits_{\lambda \lambda '} {\sum\limits_{rr'} {t_{\lambda \lambda '} \left( {\vec R_{ij} } \right)\gamma _{\lambda \sigma }^* \left( r \right)\gamma _{\lambda '\sigma }^{} \left( {r'} \right)} } },
\nonumber
\end{eqnarray}
\end{widetext}

with the hopping matrix:
\begin{widetext}
\begin{equation}
t_{\lambda \lambda '} \left( {\vec R_{ij}} \right) = \left( {\begin{array}{*{20}c}
   0 & 0 & { - 2t_{pd} \mu _{ij} } & 0 & 0  \\
   0 & 0 & {{\raise0.7ex\hbox{${2t_{pd} \xi _{ij} }$} \!\mathord{\left/
 {\vphantom {{2t_{pd} \xi _{ij} } {\sqrt 3 }}}\right.\kern-\nulldelimiterspace}
\!\lower0.7ex\hbox{${\sqrt 3 }$}}} & {{\raise0.7ex\hbox{${2t_{pd} \lambda _{ij} }$} \!\mathord{\left/
 {\vphantom {{2t_{pd} \lambda _{ij} } {\sqrt 3 }}}\right.\kern-\nulldelimiterspace}
\!\lower0.7ex\hbox{${\sqrt 3 }$}}} & 0  \\
   { - 2t_{pd} \mu _{ij} } & {{\raise0.7ex\hbox{${2t_{pd} \xi _{ij} }$} \!\mathord{\left/
 {\vphantom {{2t_{pd} \xi _{ij} } {\sqrt 3 }}}\right.\kern-\nulldelimiterspace}
\!\lower0.7ex\hbox{${\sqrt 3 }$}}} & { - 2t_{pp} \nu _{ij} } & {2t_{pp} \chi _{ij} } & { - 2t'_{pp} \xi _{ij} }  \\
   0 & {{\raise0.7ex\hbox{${2t_{pd} \lambda _{ij} }$} \!\mathord{\left/
 {\vphantom {{2t_{pd} \lambda _{ij} } {\sqrt 3 }}}\right.\kern-\nulldelimiterspace}
\!\lower0.7ex\hbox{${\sqrt 3 }$}}} & {2t_{pp} \chi _{ij} } & {2t_{pp} \nu _{ij} } & { - 2t'_{pp} \lambda _{ij} }  \\
   0 & 0 & { - 2t'_{pp} \xi _{ij} } & { - 2t'_{pp} \lambda _{ij} } & 0  \\
\end{array}} \right)
\label{eq:A.13}
\end{equation}
\end{widetext}
based on a set of the initial five orbitals, and the intracell spectrum $E_h$ in the Hamiltonian(\ref{eq:A.12}) takes  shown in Fig.(\ref{fig:2}a), where the index $h$ runs over all possible two-hole states in the $N_ -  $ sector  for the material with  magnetic ions  Cu$^{2+}$ in the $d^9$ electron configuration.
Here $\gamma _{\lambda \sigma } \left( r \right) = \left\langle {h|\rho_{\lambda \sigma }^ +  |b^s_{1g} } \right\rangle$ for any one-hole operator $\rho_{\lambda \sigma }$ (see Eq.(A.9)) and  $X_i^{ + r}  = \left| h \right\rangle \left\langle {b^s_{1g} } \right|\left( {\left| {{}^1A} \right\rangle \left\langle {b^s_{1g} } \right|} \right)$ is the Hubbard operators of creating  holes(electrons)  with the root vectors $r = \left( {h,b^s_{1g} } \right)$.

\section{The superexchange interaction in symmetric cell representation}

The superexchange interaction appears in the second order of the cell perturbation theory with respect to the hopping processes $\hat H_{cc}$  in the Hamiltonian(\ref{eq:A.12}), which corresponds to virtual excitations through the dielectric gap into the conduction band and back to valence band. These quasiparticle excitations correspond to the virtual electron-hole pairs and are described by off-diagonal elements with root vectors  $r = (h,b_{1g}^{s}) $ and $(b_{1g}^{s},^1A)$, where $s=\pm 1/2$. To highlight these contributions, we use a set of projection operators ${\hat{p}_h}$  and ${\hat{p}_0}$ , that generalized the Hubbard model analysis~\cite{Chao_etal1977} in the Mott-Hubbard approach with an arbitrary quasiparticle spectrum~\cite{Gavrichkov2016}, where
\begin{equation}
{\hat{p}_0} = \left( {X_i^{^1A^1A} + \sum\limits_{\sigma} {X_i^{b_{1g}^{\sigma}{b_{1g}^{\sigma}}} }} \right)\left( {X_j^{^1A^1A} + \sum\limits_{\sigma'} {X_j^{b_{1g}^{\sigma'}b_{1g}^{\sigma'}}} } \right)
\label{eq:B.1}
\end{equation}
and
\begin{equation}
{\hat{p}_h} = X_i^{hh} + X_j^{hh} - X_i^{hh}\sum\limits_{h'} {X_j^{h'h'}}
\label{eq:B.2}
\end{equation}

where the index $h$ runs over all $N_h=N_{nS}+N_{mT}$ two-hole states in the Fig.(\ref{fig:2}a). These operators satisfies the relations $\sum\limits_{h = 1}^{{N_h}} {{\hat{p}_h}}+ \hat{p}_0= 1$, ${\hat{p}_h}{\hat{p}_{h'}} = {\delta _{hh'}}{\hat{p}_h}$ and ${\hat{p}_h}{\hat{p}_{0}} = 0$.
We introduce the Hamiltonian of the exchange coupled ($i, j$) -pairs: ${\hat h_{ij}} = \left( {{{\hat h}^0_{ij}} + \hat h_{ij}^{in}} \right) + \hat h_{ij}^{out}$, where  $\left( {{{\hat h}^0_{ij}} + \hat h_1^{in}} \right) = {{\hat{p}_0}{{\hat h}_{ij}}{\hat{p}_{0}}}+\sum\limits_{hh'} {{\hat{p}_h}{{\hat h}_{ij}}{\hat{p}_{h'}}}$  and  $\hat h_1^{out} = \left( {\sum\limits_h {{\hat{p}_h}} } \right){\hat h_{ij}}\hat{p}_0 +  \hat{p}_0 {\hat h_{ij}}\left( {\sum\limits_h {{\hat{p}_h}} } \right)$ are the intra-cell and intra-, interband contributions for intercell path of $\hat H_{cc} = \sum\limits_{ij} {{{\hat h}_{ij}}} $ respectively, where ${\hat h}_{ij}={\hat h_{ij}^{\left( b \right)}  + \hat h_{ij}^{\left( a \right)}  + \hat h_{ij}^{\left( {ab} \right)} } $. In the unitary transformation the Hamiltonian for  $(i,j) - th$ pairs is equal to ${\hat h_{ij}} = {e^{\hat G}}{\hat h_{ij}}{e^{ - \hat G}}$ , where $\hat G$  satisfies the equation
\begin{eqnarray}
\left( {\sum\limits_h {{\hat{p}_h}} } \right){{\hat h}_{ij}}\hat{p}_0  &+& \hat{ p_0} {{\hat h}_{ij}}\left( {\sum\limits_h {{\hat{p}_h}} } \right) +
\nonumber \\
&+& \left[ {\hat G,\left( {\sum\limits_{hh'} {{\hat{p}_h}{{\hat h}_{ij}}{\hat{p}_{h'}}}  + {\hat{p}_0}{{\hat h}_{ij}}{\hat{p}_{0}}}  \right)} \right] = 0, \nonumber \\
 \label{eq:B.3}
 \end{eqnarray}
and the transformed Hamiltonian $\hat{h}_{ij}$ in the second order
of cell perturbation theory over interband hopping $\hat{h}^{out}_{ij}$ can be represented as
\begin{widetext}
\begin{equation}
\hat h_{ij} \approx \left( {\sum\limits_{hh'} {{\hat{p}_h}} {{\hat h}_{ij}}{\hat{p}_{h'}} + {{\hat{p}_0}{{\hat h}_{ij}}{\hat{p}_{0}}} } \right) + \frac{1}{2}\left[ {\hat G,\left\{ {\left( {\sum\limits_h {{\hat{p}_h}} } \right){{\hat h}_{ij}}{{\hat{p}_0}}  +  {{\hat{p}_0}} } {{\hat h}_{ij}}\left( {\sum\limits_h {{\hat{p}_h}} } \right) \right\}} \right]
\label{eq:B.4}
\end{equation}
\end{widetext}
 where
\begin{eqnarray}
&&\left( {\sum\limits_h {{\hat{p}_h}} } \right){\hat h_{ij}}{{\hat{p}_0}} =
\sum\limits_{h\sigma } {\sum\limits_{\lambda \lambda'} {t_{\lambda \lambda'} v _{i\lambda\sigma }^ +\left( hb^{\sigma}_{1g} \right)} } c _{j\lambda'\sigma }\left( {^1Ab^{\sigma}_{1g}} \right),
\nonumber \\
&&{\hat{p}_0} {\hat h_{ij}}\left( {\sum\limits_h {{\hat{p}_h}} } \right) = \sum\limits_{h\sigma } {\sum\limits_{\lambda\lambda'} {t_{\lambda \lambda'}} c_{i\lambda\sigma }^ + \left( {^1Ab^{\sigma}_{1g}}\right)v _{j\lambda'\sigma }\left(hb^{\sigma}_{1g} \right)}  \nonumber \\
\label{eq:B.5}
\end{eqnarray}
where $t_{\lambda \lambda'}$ are given in the Eq.(\ref{eq:A.12})  and
 \begin{eqnarray}
\hat G = \sum\limits_{h\sigma}&&{\sum\limits_{\lambda\lambda'}\frac{t_{\lambda\lambda'}(\vec{R_{ij}})}{\Delta _{^1Ab_{1g}h}}\left[ {  {v _{i\lambda\sigma }^ + \left(hb^{\sigma}_{1g} \right)c_{j\lambda'\sigma }^{}\left( {{^1Ab^{\sigma}_{1g}}}\right)}  - } \right.}
\nonumber \\
&&\left. { -  {c _{i\lambda\sigma }^ + \left( {^1Ab^{\sigma}_{1g}} \right)v_{j\lambda'\sigma }^{}\left( hb^{\sigma}_{1g} \right)} } \right]
\label{eq:B.6}
\end{eqnarray}
 with the energy denominator $\Delta _{^1Ab_{1g}h} = \left( {{\varepsilon _h} + {\varepsilon _{^1A}}} \right) - 2{\varepsilon _{b_{1g}}}$ and operators $c_{i \lambda \sigma}$ and $v_{i \lambda \sigma}$:

 \begin{eqnarray}
 &&c_{i \lambda \sigma}(^1A,b^{\sigma}_{1g})= \gamma _{\lambda \sigma } (^1A,b^{\sigma}_{1g})X_f^{^1A,b^{\sigma}_{1g}},  \label{eq:B.7} \\
 &&v_{i \lambda \sigma}(b^{\sigma}_{1g},nS)=  {\gamma _{\lambda \sigma } (b^{- \sigma}_{1g},\;nS)X_i^{ b^{- \sigma}_{1g},\;nS} } \nonumber \\
 &&v_{i \lambda \sigma}(b^{\sigma}_{1g},mT|_{=2\sigma})={\gamma _{\lambda \sigma } (b^{\sigma}_{1g},m2\sigma)} {X_f^{b^{\sigma}_{1g},\;m2\sigma }    } \nonumber\\
 &&v_{i \lambda \sigma}(b^{\sigma}_{1g},mT|_{=0})={\gamma _{\lambda \sigma } (b^{\sigma}_{1g},m0)}\frac{1}{{\sqrt 2 }}X_f^{b^{-\sigma}_{1g} ,\;m0}\nonumber
 \end{eqnarray}
 where $\rho_{i\lambda\sigma}=c_{i \lambda \sigma}(^1A,b^{\sigma}_{1g})+\sum\limits_h {v _{i\lambda \sigma } (b^{\sigma}_{1g},h)}$ in accordance with the Eq.(\ref{eq:A.9}). For the singlet and triplet bands at the ($b^{- \sigma}_{1g},\;nS)$ and $(b^{\sigma}_{1g},\;mT)$  root vectors,  the commutator  $\left[ {\hat{p}_0 H\hat{p}_h  ,\;\;\hat{p}_h  H\hat{p}_0 } \right]$ is determined by the operators:
\begin{widetext}
\begin{eqnarray}
&&\sum\limits_{iji'j'} {} \sum\limits_{\sigma \sigma '} {} \left[ {X_i^{b^{\sigma}_{1g}, ^1A} X_j^{ - b^{\sigma}_{1g},nS} ,\;X_{i'}^{nS, b^{- \sigma '}_{1g}} X_{j'}^{^1Ab^{\sigma'}_{1g}} } \right], \label{eq:B.8} \\
&&\sum\limits_{iji'j'} {} \sum\limits_{\sigma \sigma '}\left[ {X_i^{b^{\sigma}_{1g},^1A} \left( {X_j^{b^{\sigma}_{1g}, m2\sigma }  + {\textstyle{1 \over {\sqrt 2 }}}X_j^{ b^{- \sigma}_{1g} ,m0} } \right)\left( {X_{i'}^{m2\sigma ',b^{\sigma'}_{1g}}  + {\textstyle{1 \over {\sqrt 2 }}}X_{i'}^{m0, b^{- \sigma'}_{1g}} } \right)X_{j'}^{^1A,b^{\sigma'}_{1g}} } \right] \nonumber
\end{eqnarray}
\end{widetext}

with the AFM and FM exchange contributions respectively to the spin Hamiltonian $\hat{H}_S$ in the Eq.(\ref{eq:1}).
\bibliography{my}

\providecommand{\noopsort}[1]{}\providecommand{\singleletter}[1]{#1}%
\begin{thebibliography}{10}

\bibitem{Annet_etal1989}
J.~F. Annett, R.~M. Martin, A.~K. McMahan, and S.~Satpathy.
\newblock Electronic hamiltonian and antiferromagnetic interactions in
  {L}a$_2${C}u{O}$_4$,
\newblock {Phys. Rev. B \textbf{40}}, 2620 (1989).

\bibitem{Kim1989}
Y.~J. Kim, A.~Aharony, R.~J. Birgeneau, F.~C. Chou, O.~Entin-Wohlman, R.~W.
  Erwin, M.~Greven, A.~B. Harris, M.~A. Kastner, I.~Ya. Korenblit, Y.~S. Lee,
  and G.~Shirane.
\newblock Ordering due to quantum fluctuations in
  {S}r$_2${C}u$_3${O}$_4${C}l$_2$,
\newblock {Phys. Rev. Lett. \textbf{83}(4)}, 852 (1999).

\bibitem{Coldea2001}
R.~Coldea, S.~M. Hayden, G.~Aeppli, T.~G. Perring, C.~D. Frost, T.~E. Mason,
  S.~W. Cheong, and Z.~Fisk.
\newblock Spin waves and electronic interactions in {L}a$_2${C}u{O}$_4$.
\newblock {Phys. Rev. Lett. \textbf{86}(23)}, 5377 (2001).

\bibitem{Katanin_2002}
A~A Katanin and A~P Kampf.
\newblock Spin excitations in {L}a$_2${C}u{O}$_4$: {C}onsistent description by
  inclusion of ring exchange,
\newblock {Phys. Rev. B \textbf{66}}, 100403(R) (2002).

\bibitem{Toader_2005}
A.~M. Toader, J.~P. Goff, M.~Roger, N.~Shannon, J.~R. Stewart, and M.~Enderle.
\newblock Spin correlations in the paramagnetic phase and ring exchange in
  {L}a$_2${C}u{O}$_4$,
\newblock {Phys. Rev. Lett. \textbf{94}}, 197202 (2005).

\bibitem{Headings_2010}
N.~S. Headings, S.~M. Hayden, R.~Coldea, and T.~G. Perring.
\newblock Anomalous high-energy spin excitations in the high-{T}$_c$
  superconductor-parent antiferromagnet {L}a$_2${C}u{O}$_4$,
\newblock {Phys. Rev. Lett. \textbf{105}}, 247001 (2010).

\bibitem{Dean_2012}
M.~P.~M. Dean, R.~S. Springell, C.~Monney, K.~J. Zhou, J.~Pereiro, I.~Bozovic,
  D.~B. Piazza, H.~M. Ronnow, E.~Morenzoni, J.~van~den Brink, T.~Schmitt, and
  J.~P. Hill.
\newblock Spin excitations in a single {L}a$_2${C}u{O}$_4$ layer,
\newblock {Nature Materials \textbf{11}}, 850 (2012).

\bibitem{Yamamoto_2019}
S.~Yamamoto and Y.~Noriki.
\newblock Spin-wave thermodynamics of square-lattice antiferromagnets
  revisited,
\newblock {Phys. Rev.B \textbf{99}}, 094412 (2019).

\bibitem{Bao_2025}
J.~Bao, M.~Gohlke, J.~G. Rau, and N.~Shannon.
\newblock Magnon spectra of cuprates beyond spin wave theory,
\newblock {Phys. Rev. Research \textbf{7}}, L012053 (2025).

\bibitem{Qin2020}
M.~Qin.
\newblock Absence of superconductivity in the pure two-dimensional {H}ubbard
  model,
\newblock {Phys. Rev. X \textbf{10}}, 031016 (2020).

\bibitem{Xu_2024}
H.~Xu, C.-M. Chung, M.~Qin, U.~Schollwock, S.~R. White, and S.~Zhang.
\newblock Coexistence of superconductivity with partially filled stripes in the
  {H}ubbard model,
\newblock {Science \textbf{384}}, 7691 (2024).

\bibitem{Feiner_etal1996}
L.~F. Feiner, J.~H. Jefferson, and R.~Raimondi.
\newblock Effective single band models for high {T}$_c$ cuprates. {I}.
  {C}oulomb interactions,
\newblock {Phys. Rev. B \textbf{53}}, 8751 (1996).

\bibitem{Raimondy_etal1996}
R.~Raimondi, J.~H. Jefferson, and L.~F. Feiner.
\newblock Effective single-band models for the high-{T}$_c$ cuprates.{ II}.
  {R}ole of apical oxygen,
\newblock {Phys. Rev. B \textbf{53}}, 8774 (1996).

\bibitem{Gavrichkov_2000}
V.~A. Gavrichkov, S.~G. Ovchinnikov, A.~A. Borisov, and E.~G. Goryachev.
\newblock Evolution of the band structure of quasiparticles with doping in
  copper oxides on the basis of a generalized tight-binding method.
\newblock {JETP \textbf{91}(2)}, 369 (2000).

\bibitem{Gavrichkov2017}
V.~A. Gavrichkov, S.~I. Polukeev, and S.~G. Ovchinnikov.
\newblock Contribution from optically excited many-electron states to the
  superexchange interaction in {M}ott-{H}ubbard insulators,
\newblock {Phys. Rev. B \textbf{95}}, 144424 (2017).

\bibitem{Gavrichkov2016}
V.~A. Gavrichkov, Z.~V. Pchelkina, I.~A. Nekrasov, and S.~G. Ovchinnikov.
\newblock Pressure effect on the energy structure and superexchange interaction
  in undoped orthorhombic {L}a$_2${C}u{O}$_4$,
\newblock {International Journal of Modern Physics B \textbf{30}}, 1650180 (2016).

\bibitem{Gavrichkov2020}
V.~A. Gavrichkov, S.~I. Polukeev, and S.~G. Ovchinnikov.
\newblock Cation spin and superexchange interaction in oxide materials below
  and above spin crossover under high pressure,
\newblock {Phys. Rev. B \textbf{101}}, 094409 (2020).

\bibitem{Mikhaylovskiy2020}
R.~V. Mikhaylovskiy, T.~J. Huisman, V.~A. Gavrichkov, S.~I. Polukeev, S.~G.
  Ovchinnikov, D.~Afanasiev, R.~V. Pisarev, Th. Rasing, and A.~V. Kimel.
\newblock Resonant pumping of d-d crystal field electronic transitions as a
  mechanism of ultrafast optical control of the exchange interactions in iron
  oxides,
\newblock {Phys. Rev. Lett. \textbf{125}}, 157201 (2020).

\bibitem{Jiang2019}
H.~C. Jiang and T.~P. Devereaux.
\newblock Superconductivity in the doped {H}ubbard model and its interplay with
  next-nearest hopping $t'$,
\newblock {Science \textbf{365}}, 1424 (2019).

\bibitem{White1999}
S.~R. White and D.~J. Scalapino.
\newblock Competition between stripes and pairing in a $t$-$t_0$-{J} model,
\newblock {Phys. Rev. B \textbf{60}}, R753 (1999).

\bibitem{Chung2020}
C.-M. Chung, M.~Qin, S.~Zhang, U.~Schollwock, and S.~R. White.
\newblock Plaquette versus ordinary d-wave pairing in the $t'$-{H}ubbard model
  on a width 4 cylinder,
\newblock {Phys. Rev. B \textbf{102}}, 041106(R) (2020).

\bibitem{Wei2024}
W.~He, J.~Wen, H.-C. Jiang, G.~Xu, W.~Tian, T.~Taniguchi, Y.~Ikeda, M.~Fujita,
  and Y.~S. Lee.
\newblock Tilted stripes origin in {L}a$_{1.88}${S}r$_{0.12}${C}u{O}$_4$
  revealed by anisotropic next-nearest,
\newblock {Commun Phys \textbf{7}}, 257 (2024).

\bibitem{Jiang2022}
S.~Jiang, D.~J. Scalapino, and S.~R. White.
\newblock Pairing properties of the $t$-$t'$ -$t"$ -{J} model,
\newblock {Phys. Rev. B \textbf{106}}, 174507 (2022).

\bibitem{Hartstein2020a}
M.~Hartstein, Yu.~Te. Hsu, K.~A. Modic, J.~Porras, T.~Loew, M.~LeTacon, H.~Zuo,
  J.~Wang, Z.~Zhu, M.~K. Chan, R.~D. McDonald, G.~G. Lonzarich, B.~Keimer,
  S.~E. Sebastian, and N.~Harrison.
\newblock Hard antinodal gap revealed by quantum oscillations in the pseudogap
  regime of underdoped high-{T}$_c$ superconductors,
\newblock {Nature Physics \textbf{16}}, 841 (2020).

\bibitem{Kanigel2006}
A.~Kanigel, M.~R. Norman, M.~Randeria, U.~Chatterjee, S.~Souma, A.~Kaminski,
  H.~M. Fretwell, S.~Rozenkranz, M.~Shi, T.~Sato, T.~Takahashi, Z.~Z. Li,
  H.~Raffy, K.~Kadowaki, D.~Hinks, L.~Ozyuzer, and Campuzano~J. C.
\newblock Evolution of the pseudogap from {F}ermi arcs to the nodal liquid,
\newblock {Nature Physics \textbf{2}}, 447 (2006).

\bibitem{AlRashid2024}
H.~Al-Rashid and D.~K. Singh.
\newblock Effect of next-nearest neighbor hopping on the single-particle
  excitations at finite temperature.
\newblock {SciPost Phys. \textbf{16}}, 107 (2024).

\bibitem{Chen2012}
K.~S. Chen, Z.~Y. Meng, T.~Pruschke, J.~Moreno, and M.~Jarrell.
\newblock Lifshitz transition in the two-dimensional {H}ubbard model,
\newblock {Phys. Rev. B \textbf{86}}, 165136 (2012).

\bibitem{Kohno2014}
M.~Kohno.
\newblock Spectral properties near the {M}ott transition in the two-dimensional
  hubbard model with next-nearest-neighbor hopping,
\newblock {Phys. Rev. B \textbf{90}}, 035111 (2014).

\bibitem{Gull2009}
E.~Gull, O.~Parcollet, P.~Werner, and A.~J. Millis.
\newblock Momentum-sector-selective metal-insulator transition in the
  eight-site dynamical mean-field approximation to the {H}ubbard model in two
  dimensions,
\newblock {Phys. Rev. B \textbf{80}}, 245102 (2009).

\bibitem{Wakimoto1999}
S.~Wakimoto, G.~Shirane, Y.~Endoh, K.~Hirota, S.~Ueki, K.~Yamada, R.~J.
  Birgeneau, M.~A. Kastner, Y.~S. Lee, P.~M. Gehring, and S.~H. Lee.
\newblock Observation of incommensurate magnetic correlations at the lower
  critical concentration for superconductivity in
  {L}a$_{2-x}${S}r$_x${C}u{O}$_4$ $(x=0.05)$,
\newblock {Phys. Rev. B \textbf{60}}, R769 (1999).

\bibitem{Fujita2002}
M.~Fujita.
\newblock Static magnetic correlations near the insulating superconducting
  phase boundary in {L}a$_{2-x}${S}r$_x${C}u{O}$_4$,
\newblock {Phys. Rev. B \textbf{65}}, 064505 (2002).

\bibitem{Dunsiger2008}
S.~R. Dunsiger, Y.~Zhao, B.~D. Gaulin, Y.~Qiu, P.~Bourges, Y.~Sidis, J.~R.~D.
  Copley, A.~Kallin, E.~M. Mazurek, and H.~A. Dabkowska.
\newblock Diagonal and collinear incommensurate spin structures in underdoped
  {L}a$_{2-x}${B}a$_x${C}u{O}$_4$,
\newblock {Phys. Rev. B \textbf{78}}, 092507 (2008).

\bibitem{Moreira2006}
I.~P.~R. Moreira, C.~J. Calzado, J.-P. Malrieu, and F.~Illas.
\newblock First-principles periodic calculation of four-body spin terms in
  high-{T}$_c$ cuprate superconductors
\newblock {Phys. Rev. Lett. \textbf{97}}, 087003 (2006).

\bibitem{Wan2009}
X.~Wan, T.~A. Maier, and S.~Y. Savrasov.
\newblock Calculated magnetic exchange interactions in high-temperature
  superconductors,
\newblock {Phys Rev B \textbf{79}}, 155114 (2009).

\bibitem{Kamimura_etal1990}
H.~Kamimura and M.~Eto.
\newblock $^1${A}$_{1g}$ to $^3${B}$_{1g}$ conversion at the onset of
  superconductivity in {L}a$_{2-x}${S}r$_x${C}u{O}$_4$ due to the apical oxygen
  effect,
\newblock {J. Phys. Soc. Jpn \textbf{59}}, 3053 (1990).

\bibitem{Eto1991}
M.~Eto and H.~Kamimura.
\newblock Low spin to high spin transition at the onset of superconductivity in
  {L}a{S}r{C}uo and high spin state in {N}d{C}e{C}u{O} compound,
\newblock {Physica C Superconductivity \textbf{185-189}}, 1599 (1991).

\bibitem{Janowitz2004}
C.~Janowitz, U.~Seidel, R.-S.~T. Unger, A.~Krapf, R.~Manzke, V.~Gavrichkov, and
  S.~Ovchinnikov.
\newblock Strong spin triplet contribution of the first removal state in the
  insulating regime of {B}i$_2${S}r$_2${C}a$_{1- x}${Y}$_x${C}u$_2${O}$_{8+y}$,
\newblock {JETP Letters \textbf{80}(11)}, 692 (2004).

\bibitem{Anderson1959}
P.~W. Anderson.
\newblock New approach to the theory of superexchange interactions,
\newblock {Phys. Rev. \textbf{115}}, 2 (1959).

\bibitem{Chao_etal1977}
K.~A. Chao, J.~Spalek, and A.~M. Oles.
\newblock Kinetic exchange interaction in a narrow {S}-band,
\newblock {Journal of Physics C: Solid State Physics \textbf{10}}, L271 (1977).

\bibitem{Irkhin1994}
V.~Y. Irkhin and Y.~P. Irkhin.
\newblock Many-electron operator approach in the solid state theory,
\newblock {Phys. Status Solidi B \textbf{183}}, 9 (1994).

\bibitem{Hubbard_1963}
J.~Hubbard.
\newblock Electron correlation in narrow band,
\newblock {Proc. Roy. Soc. \textbf{A276}}, 238 (1963).

\bibitem{Shastry_1989}
B.~S. Shastry.
\newblock $t$-{J} model and nuclear magnetic relaxation in high-${T}_c$
  materials,
\newblock {Phys. Rev. Lett. \textbf{63}}, 1288 (1989).

\bibitem{Feiner_etalPRL1996}
L.~F. Feiner, J.~H. Jefferson, and R.~Raimondi.
\newblock Intrasublattice hopping in the extended $t$-{J} model and
  {T}$_{cmax}$ in the cuprates,
\newblock {Phys. Rev. Lett. \textbf{76}}, 4939 (1996).

\bibitem{Spalek2007}
J.~Spalek.
\newblock $t$-{J} model then and now: {A} personal perspective from the
  pioneering times,
\newblock {Acta Phys. Pol. A \textbf{111}(4)}, 409 (2007).

\bibitem{Jefferson1990}
J.H. Jefferson.
\newblock Derivation of the $t$-{J} model for high temperature
  superconductivity,
\newblock {Physica B: Condensed Matter \textbf{165-166}(P2)}, 1013 (1990).

\bibitem{Chao1977}
K.~A. Chao, J.~Spalek, and A.~M. Oles.
\newblock The kinetic exchange interaction in doubly degenerate narrow bands.
\newblock {Phys. Stat. Sol. (b) \textbf{84}}, 747 (1977).

\bibitem{Spalek1980}
J.~Spalek and K.~A. Chao.
\newblock Kinetic exchange interaction in a doubly degenerate narrow band and
  its application to {F}e$_{1- x}${C}o$_x${S}$_2$ and
  {C}o$_{1-x}${N}i$_x${S}$_2$,
\newblock {J. Phys. C: Solid State Phys. \textbf{13}}, 5241 (1980).

\bibitem{Bianconi_etal1996}
A.~Bianconi, N.~L. Saini, A.~Lanzara, M.~Missori, T.~Rossetti, H.~Oyanagi,
  H.~Yamaguchi, K.~Oka, and T.~Ito.
\newblock Determination of the local lattice distortions in the {C}u{O}$_2$
  plane of {L}a$_{1.85}${S}r$_{0.15}${C}u{O}$_4$,
\newblock {Phys. Rev. Lett. \textbf{76}}, 3412 (1996).

\bibitem{Bianconi_2000}
A.~Bianconi, N.~L. Saini, S.~Agrestini, D.~Di~Castro, and G.~Bianconi.
\newblock The strain quantum critical point for superstripes in the phase
  diagram of all cuprate perovskites,
\newblock {International Journal of Modern Physics B \textbf{14}}, 3342 (2000).

\bibitem{Albertini_2023}
R.~Albertini, S.~Macis, A.~A. Ivanov, A.~P. Menushenkov, A.~Puri,
  V.~Monteseguro, B.~Joseph, W.~Xu, A.~Marcelli, P.~Giraldo-Gallo, I.~R.
  Fisher, A.~Bianconi, and G.~Campi,
\newblock Tensile microstrain fluctuations in the {B}a{P}b{O} units in
  superconducting {B}a{P}b$_{1-x}${B}i$_x${O}$_3$ by scanning dispersive
  micro-{X}{A}{N}{E}{S},
\newblock {Condensed Matter \textbf{8}(3)}, 57 (2023).

\bibitem{Kugel_2008}
K.~I. Kugel, A.~L. Rakhmanov, A.~O. Sboychakov, N.~Poccia, and A.~Bianconi.
\newblock Model for phase separation controlled by doping and the internal
  chemical pressure in different cuprate superconductors.
\newblock {Phys. Rev. B \textbf{78}(16)}, 165124 (2008).

\bibitem{Sboychakov_2022}
A.~O. Sboychakov, K.~I. Kugel, and A.~Bianconi.
\newblock Moire-like superlattice generated van hove singularities in a
  strained {C}uo$_2$ double layer,
\newblock {Condensed Matter \textbf{7}(3)}, 50 (2022).

\bibitem{Vinograd_2024}
I.~Vinograd, S.~M. Souliou, A.~A. Haghighirad, T.~Lacmann, Y.~Caplan,
  M.~Frachet, M.~Merz, G.~Garbarino, Y.~Liu, S.~Nakata, K.~Ishida, H.~M.~L.
  Noad, M.~Minola, B.~Keimer, D.~Orgad, C.~W. Hicks, and M.~Le~Tacon.
\newblock Using strain to uncover the interplay between two- and
  three-dimensional charge density waves in high-temperature superconducting
  {Y}{B}a$_2${C}u$_3${O}$_y$,
\newblock {Nat. Commun. \textbf{15}}, 3277 (2024).

\bibitem{Hameed_2024}
S.~Hameed, Y.~Liu, K~S Rabinovich, G.~Kim, P.~Wochner, G.~Christiani,
  G.~Logvenov, K.~Higuchi, N.~B. Brookes, E.~Weschke, F.~Yakhou-Harris, A.~V.
  Boris, B.~Keimer, and M.~Minola.
\newblock Interplay between electronic and lattice superstructures in
  {L}a$_{2-x}${C}a$_x${C}u{O}$_4$,
\newblock {arXiv: \textbf{2408.06774}}, (2024).

\bibitem{Valkov1982}
V.V. Val'kov and S.G. Ovchinnikov.
\newblock Hubbard operators and spin-wave theory of {H}eisenberg magnets with
  arbitrary spin.
\newblock {Teoreticheskaya i Matematicheskaya Fizika \textbf{50}(2)}, 466 (1982).

\bibitem{Downey2023}
P.-O Downey, O.~Gingras, J.~Fournier, C.-D. Hebert, M.~Charlebois, and A.-M.~S.
  Tremblay.
\newblock Mott transition, {W}idom line, and pseudogap in the half-filled
  triangular lattice {H}ubbard model
\newblock {Phys. Rev. B \textbf{107}}, 125159 (2023).

\bibitem{Korshunov_etal2005}
M.~M. Korshunov, V.~A. Gavrichkov, S.~G. Ovchinnikov, I.~A. Nekrasov, Z.~V.
  Pchelkina, and V.~I. Anisimov.
\newblock Hybrid {L}{D}{A} and generalized tight-binding method for electronic
  structure calculations of strongly correlated electron systems,
\newblock {Phys. Rev. B \textbf{72}}, 165104 (2005).

\bibitem{Pavarini_etal2001}
E.~Pavarini, I.~Dasgupta, T.~Saha-Dasgupta, O.~Jepsen, and O.~K. Andersen.
\newblock Band-structure trend in hole-doped cuprates and correlation with
  {T}$_{cmax}$,
\newblock {Phys. Rev. Lett. \textbf{87}}, 047003 (2001).

\bibitem{Zaitsev_1976}
R.~O. Zaitsev.
\newblock Diagram technique and gas approximation in the {H}ubbard model,
\newblock {JETP \textbf{43}}, 574 (1976).

\bibitem{Ovchinnikov_etal2004}
S.~G. Ovchinnikov and V.~V. Val'kov.
\newblock {Hubbard operators in the theory strongly correlated electrons},
\newblock Imperial College Press, London (2004).

\bibitem{Ovchinnikov_etal2012}
S.~G. Ovchinnikov, V.~A. Gavrichkov, M.~M. Korshunov, and E.~I. Shneyder.
\newblock {Springer Series in Solid-State Sciences, {T}heoretical {M}ethods for strongly {C}orrelated systems}, 
\textbf{171}, chapter: LDA+GTB method for band structure calculations in the strongly correlated
materials, pp.143--171 (2012).

\end{thebibliography}
\end{document}